\documentclass{aa}  
\usepackage{txfonts}

\usepackage{newtxtext,newtxmath}

\usepackage[T1]{fontenc}
\usepackage{ae,aecompl}

\usepackage{graphicx}	
\usepackage{amsmath}	
\usepackage{amssymb}	
\usepackage{rotating}
\usepackage{listings}
\usepackage{threeparttable}

\begin{document} 
            \title{X-ray absorption in INTEGRAL AGN:}
             
             \subtitle{Host galaxy inclination}

\author{A. Malizia, \inst{1}
L. Bassani,  \inst{1}
J. B. Stephen,  \inst{1}
A. Bazzano  \inst{2}
\and P. Ubertini \inst{2}
}

\institute{OAS-IASF-Bologna, Via P. Gobetti 101, I-40129 Bologna, Italy\\
              \email{angela.malizia@inaf.it}
         \and
               IAPS-IASF-Roma, Via Fosso del Cavaliere 100, I-00133, Roma, Italy \\
}

\date{ Received ; accepeted}

\abstract 
{In this work the INTEGRAL hard X-ray selected sample of AGN has been
used to investigate the possible contribution of absorbing material distributed within the host
galaxies to the total amount of N$_{H}$  measured in the X-ray band.
We collected all the available axial
ratio measurements of the galaxies hosting our AGN together with their
morphological information and find that also for our hard X-ray selected sample a deficit of
edge-on galaxies hosting type 1 AGN is present.
We estimate that in our hard X-ray selected sample there is a deficit of 24\% ($\pm$ 5\%) of type 1 AGN. Possible bias in redshift has been
excluded, as  we found the same effect in a well determined range of z   
where the number and the distributions of the two classes are statistically the same.
Our findings clearly indicate that material located in the host galaxy on scales of hundreds
of parsecs and not aligned with the putative absorbing torus of the AGN can contribute to the
total amount of column density. This galactic absorber can be large enough to hide the broad
line region of some type 1 AGN causing their classification as type 2 objects and giving rise
to the deficiency of type 1 in edge-on galaxies.}


   \maketitle



\section{Introduction}
The nature of the absorbing material hiding the central engine of active galactic nuclei (AGN) is crucial to our understanding of 
the physics of these objects \citep{2018ARA&A..56..625H}. It is also a key issue in the unified model of AGN \citep{Antonucci_1993, Urry_1995}  
which in its simplest version postulates that the diversity of AGN can be largely explained as a viewing angle effect.  
The most important ingredient of this orientation-based model is an optically and geometrically thick  torus that obscures the 
nuclear regions of an active galaxy (the accretion disk and the hot corona  as well as the broad line region (BLR)). 
We optically classify an AGN as type 2 or type 1 depending if our line of sight intercepts or not the obscuring material of the 
torus.\\ 
The validity of the unified model of AGN has largely been tested and confirmed by the presence of high absorption 
measured in the X-ray spectra  of type 2  with respect to  lower amounts found in type 1 AGN.
In particular, using  a  sample of 272 AGN observed at high energies ($>$ 20 keV) by INTEGRAL/IBIS we have found that the 
standard-based AGN unification scheme is followed by the majority  of bright AGN \citep{Malizia_2012} with only a few exceptions
related to 12--13\% of the objects.
These exceptions are absorbed type 1 and unabsorbed type 2 AGN. The absorption in type 1 AGN is generally  interpreted in terms 
of ionised gas located in an accretion disc wind or in the biconical structure associated with the central nucleus 
and therefore unrelated to the torus. The lack of  X-ray absorption (N$_{H}<$ 10$^{22}$ cm$^{-2}$) in type 2 AGN could 
be explained by the assumption that the torus is either not present (or has disappeared) or the source has been misclassified. 
The first case is probably true for  low luminosity objects \citep{Elitzur_2008} while the second case is relevant for the 
class of intermediate type 2 Seyfert (type 1.8 -- 1.9).
These intermediate objects have  historically been considered as AGN where the observer's line of  sight intercepts either 
the outer edge of the torus or a limited number of clouds, so that the broad line region is still partly visible. 
However, in \citet{Malizia_2012} we have also pointed out that these intermediate classifications can  be explained by an  intrinsically 
variable ionising continuum or by  the presence of absorption/reddening unrelated to the torus. 
For example, a source that would normally appear as a type 1 Seyfert can be classified as an intermediate type if it is in a low 
optical flux state \citep{Trippe_2010} or if  its BLR is  obscured (except for the strongest H$_{\alpha}$ line) by dust related to 
large-scale structures such as bars, dust lanes and host galaxy material \citep{Malkan_1998, 2000A&A...355L..31M}.\\
Thus absorption in AGN is probably due not only to one component, the torus, but to  multiple  components on very different scales.
\citet{Bianchi_2012} identified 3 such components: the  BLR on the 0.01 pc scale, the torus on a parsec scale and absorption located 
in the host galaxy  on a scale of hundreds of parsecs.\\
Evidence  for large scale absorption comes mostly from inclination studies of the host galaxies of AGN. 
In one of the first such studies \citet{Keel_1980} found that optically-selected (mostly type 1) Seyferts tend to avoid 
edge-on host galaxies.
Using a sample of AGN selected in the soft X-ray band (0.2-3.5 keV), \citet{Simcoe_1997}
subsequently confirmed the  bias against edge-on Seyfert 1 although they were able to recover some of the edge-on AGN missed in UV
and visible surveys, resulting in  30\% incompleteness for type 1's. 
However since the soft X-ray band can also be biased in terms of absorption, a definitive test would be provided by the use of 
a hard X-ray selected sample.
A preliminary analysis performed on a Swift/BAT sample of around 80 AGN by \citet{winter_2009} showed that objects with low X-ray column 
densities were  preferentially found in galaxies with low inclination angles (face-on, $b/a >$ 0.5), while those with higher column 
densities were found in galaxies of any inclination  (edge-on and face-on, 0.1 $<$ b/a $<$ 1).
The finding that optically selected AGN samples tend to avoid edge-on systems, has recently  been confirmed and refined with much higher statistics by using the SDSS survey \citep{Lagos_2011}.\\
The first explanations for this result came from \citet{Maiolino_1995}  and \citet{2000A&A...355L..31M}.
\citet{Maiolino_1995} proposed the {\it dual absorber} model which 
foresees the existence of a 100 pc scale obscuring material of molecular gas coplanar with the galactic disc and not necessarily 
aligned with the torus. On the other hand, taking into account observational evidence, \citet{2000A&A...355L..31M} proposed a scenario where Compton thick Seyfert 2 galaxies 
are those sources observed through the torus while Compton thin/intermediate Seyfert galaxies are obscured by dust lanes at larger distances.\\
Focusing on the possible origin of the galactic absorber, a major study is that of \citet{Malkan_1998}, who detected fine-scale 
structure in the centres of nearby galaxies using HST images: these structures include dust lanes and patches, bars, rings, 
wisps, filaments and  tidal features such as warps and tails. These authors even suggested a new unified model 
(the galactic dust model) whereby the obscuration that converts an intrinsic Seyfert 1 nucleus into an apparent Seyfert 2 occurs 
in the host galaxy hundreds of parsecs from the nucleus. More recently \citet{Prieto_2014} confirmed these early results of large scale, a few hundred pc, dust filaments and diffuse dust lanes hiding the central region of some AGN. \\
Nowadays the parsec scale environment of AGN can be spatially resolved by high angular resolution observations in the infrared (IR) with VLTI and in the sub-Millimeter by ALMA 
(see e.g. \citet{2017ApJ...838L..20H, 2019A&A...623A..79C}) which draws a much more complex picture of multi-phase and multi-component regions. 
Recently \citet{2019ApJ...884..171H} has proposed a model which accounts for these new observations wherein  dusty molecular gas flows in from the
host galaxy ($\sim$100 pc) to the sub-parsec environment via a disc with small to moderate height. Due to the radiation pressure, the disc puffs up and unbinds a large amount 
of the inflowing gas from the black hole's gravitational potential allowing the creation of dusty molecular winds driven by the radiation pressure from the AGN. 
These dusty winds feed back the host galaxy with a rate increasing with the AGN luminosity (or Eddington ratio), and are therefore a mechanism to self-regulate the AGN activity providing a feedback from AGN to host galaxy. Within this picture there are multiple spatial and dynamical components which obscure the
primary emission rather than a single torus.
For local, radio quiet active galaxies, Hoenig's  model  is consistent with the structure proposed by \citet{Ricci_2017}
to explain the X-ray obscuration in AGN.
ALMA observations have revealed the presence of  circumnuclear molecular (CO) gas in many Seyfert galaxies \citep{Schinnerer_1999}, 
as well as the presence of cold gas and dust at hundreds of pc scales in many AGN host galaxies (e.g. \citealt{Garc_a_Burillo_2005,Garc_a_Burillo_2014}).\\
Finally, we note that large scale absorption may come also from non asymmetric perturbations which provide  a viable way to  channel  
gas from the outer part of a galaxy into its central regions. These  perturbations can be of two types: external, such as 
galaxy-galaxy interactions or internal, such as due to bars and their gravity torques. 
Observations indicate that an AGN becomes heavily obscured behind merger-driven gas and dust, even in the early stages of 
galaxy-galaxy interaction, when the galaxies are still well separated \citep{Kocevski_2015}.
Furthermore,  analysis of Swift/BAT AGN indicates that a large fraction (25\%) of these objects  show  disturbed morphologies or are in close 
physical pairs  compared to  matched  control  galaxies  or  optically  selected  AGN \citep{Koss_2010}. \\
In this paper we will investigate  the presence/role  of obscuration on large galactic scales and its relation with the X-ray column density,  by examining the axial ratio distribution 
 in  a well defined sample of  INTEGRAL selected AGN, listed in \citet{Malizia_2012, Malizia_2016} 
and newly reported in this work. This sample, being hard X-ray selected, is the most appropriate to carry out this type of study since it 
is  unbiased against obscured objects and therefore free of the limitation  which affects surveys at other frequencies 
(i.e. from optical to soft X-rays).

\begin{figure}
	\includegraphics[width=\columnwidth]{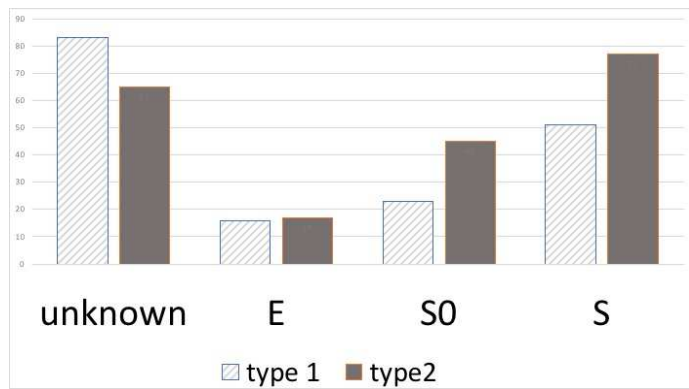}
   \caption{Histogram of the morphological classification following the \citet{DeVaucouleurs_1959} scheme. 
Unknown: galaxies which have no host galaxy morphology, E: elliptical galaxies; S0: intermediate systems between ellipticals and spirals; S: spiral galaxies. 
Objects are further sub-divided in type 1 and type 2.}
    \label{morph}
\end{figure}

\section{The sample}
In order to have a well defined hard X-ray selected sample of active galactic nuclei, we have considered all the INTEGRAL Seyferts
listed in  \citet{Malizia_2012, Malizia_2016} plus those reported in  Tables \ref{mere}.2 and \ref{krivo}.3 of the Appendix. 
This hard X-ray selected sample of AGN  has been extracted from the INTEGRAL/IBIS all sky surveys performed so far  in the 20-100 keV band (\cite{Bird_2016} and reference therein). 
This sample has been updated in this work with the addition of the new  AGN reported by 
\citet{Mereminskiy_2016}  in the  deep extragalactic surveys of M81, LMC and 3C 273/Coma regions and those reported by 
\citet{Krivonos_2017} in the Galactic Plane Survey (see Appendix and Tables \ref{mere}.2 and \ref{krivo}.3 for details on these new additions).\\
The entire INTEGRAL AGN sample is fully characterised in terms of optical identification/classification  
and is also fully studied in terms of  X-ray spectral properties (see \cite{Malizia_2012}).  
In particular, the column densities (N$_{H}$) measured in the 2-10 keV band for each AGN, have been collected from the literature (see  \citet{Malizia_2012, Malizia_2016})
or calculated for the new added sources in this work (see Appendix, Tables \ref{mere}.2 and \ref{krivo}.3).
All together, we have gathered a sample of 376 hard X-ray selected active  galaxies, with a well defined set of information. 
Objects identified as Blazars in the IBIS surveys have been excluded from the analysis to avoid complication due to the presence of jets pointing towards the observer.
Following our previous works 
we divided our AGN into type 1 comprising all the broad line Seyferts of our sample 
and including Seyfert 1, 1.2, 1.5 (179 objects) and type 2 objects comprising all the narrow line AGN i.e. Seyfert 1.8, 1.9 and 2 
(197 objects). 
Twenty-two objects ($\sim$ 6\% of the sample) have an unknown optical classification or belong to non-Seyfert optical classes (liners, XBONG, AGN), therefore they have been assimilated into type 1 or 2 AGN depending if they were absorbed (21 objects) or not in the X-ray band. Performing the complete analysis without these sources does not significantly change the overall results.\\
In order to  investigate the role of absorption in the host galaxy we have collected for the entire sample the  axial ratio $b/a$, used here as 
a proxy for the host galaxy inclination (to first order, $cos$ $i = b/a$ where a and  b are the observed major and minor axes of a galaxy and $i$ is its inclination). 
These values  have been collected from the Two Micron All Sky Survey\footnote{http://irsa.ipac.caltech.edu/cgi-bin/Gator/nph-dd}, where 
information was available for  300 objects out of  376 in the sample (i.e. the largest coverage available in catalogues with b/a informations) and they are reported in columns 5 and 10 of tables \ref{all}.1 and \ref{mere}.2/\ref{krivo}.3 respectively.
It is worth noting that  as described by \citet{2003AJ....125..525J}, the 2MASS  images, co-adding J, H and K bands, have a point spread function 
FWHM of 2 - 3 arcsec. This value is  much smaller that the radii of our galaxies (around 20-30 arcsec or more) collected from the same archive,
therefore excluding any possible bias in the $b/a$ measurements used in our work, especially for type 1 objects hosting a bright nucleus.
Furthermore, in order to verify the accuracy of these axial ratios, we have used Pan-STARRS1 database \citep{2016arXiv161205560C} and extracted the  DR1 color (g to z filter) images for a representative 
sample of over  30 of our objects ranging in axial ratio from 0.1 to 0.9 and over all optical classes. 
Representative set of objects among those with no optical classification has also been checked.
We then fitted ellipses to these images in order to measure their axial ratio values. 
In every case these values were consistent with those extracted from the 2MASS surveys and originally used. \\

\begin{figure}
	\includegraphics[width=\columnwidth]{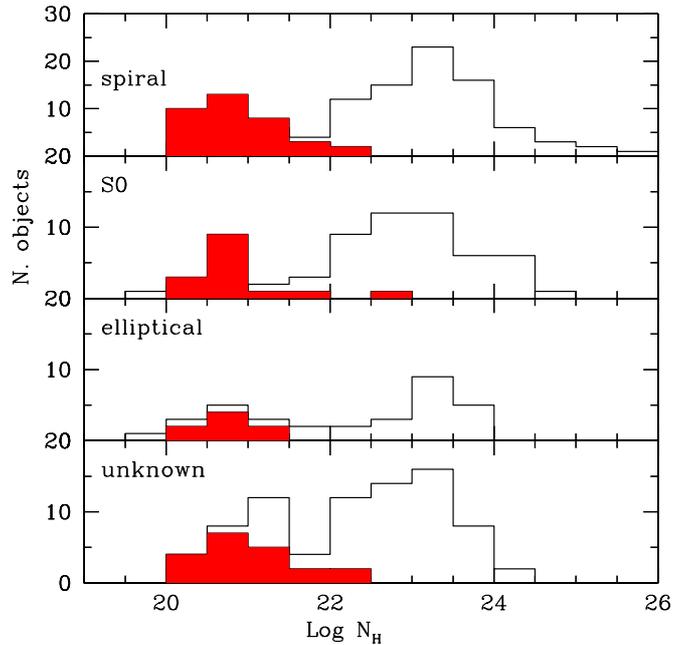}
  \caption{Distribution of column density in each morphological class; X-ray absorption in AGN hosted in galaxies of unknown class is also shown. 
  Filled red bins referred to Galactic or upper limit values of N$_{H}$.}
    \label{morph_nh}
\end{figure}

\begin{figure*}
\centering
\includegraphics[width=12cm]{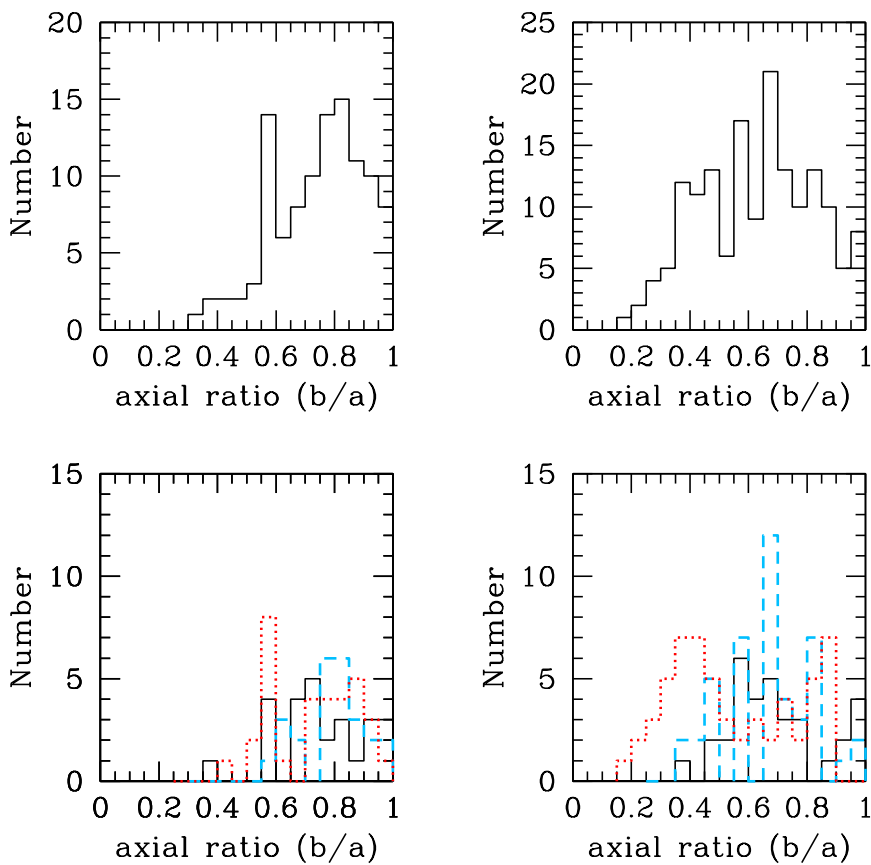}
   \caption{Distribution of host galaxy axial ratios of INTEGRAL AGN differentiated into broad lines or type 1 i.e. Sy 1 - 1.2 - 1.5 (upper and lower left panels) and narrow lines or type 2 i.e. Sy 1.8 - 1.9 - 2 (upper and lower right panels).
   In the lower panels the morphological type of our sample sources has been considered; histograms in dotted red lines  refer
    to AGN hosted in late type (Spirals) galaxies, blue dashed lines AGN hosted in early (Elliptical and S0) galaxies and black solid lines refer to host galaxies
   for which no morphological type has been found in the literature.}
    \label{histo1}
\end{figure*}

\section{Absorption and source morphology} 
Indications that absorption on a galaxy scale is important in the overall column density budget of an AGN, suggests a possible link  between 
galaxy morphology and N$_H$. 
We have therefore collected data for both these two parameters.
Data for host galaxy morphological information for all objects come from the Hyperleda catalogue  \citep{Makarov_2014}, the NED archive
and the 2MASS Redshift Survey \citep{Huchra_2012}:  the most quoted classification among these 3 databases has been adopted and reported in tables A1, A2 and A3. 
When no information or different classification were available, we have also searched the literature for extra information (\citealt{Maiolino_1999}, \citealt{Madrid_2006}, \citealt{McKernan_2010}, \citealt{Maia_2003}, \citealt{Tsvetanov_1992} and \citealt{Ferruit_2000}).
Data for N$_{H}$ values comes instead from our previous publications and current paper as specified in section 2.\\
All the INTEGRAL AGN have been  divided into  3 broad morphological types following the \citet{DeVaucouleurs_1959} scheme: 
E to describe elliptical galaxies; S0 for objects which are intermediate systems between ellipticals and spirals (alternatively called 
lenticular galaxies), and S for spiral galaxies. 
The distribution of INTEGRAL AGN in each morphological class, including those objects for which the host galaxy morphology is unknown,  is shown in figure \ref{morph}, 
where  the predominance of spiral/lenticular galaxies over ellipticals is evident.
The greater number of type 2 over  type 1 AGN in both S0 and S classes is also clear.\\
We have been able to obtain morphological host galaxy  information for  228 objects or 61\% of the sample.  
This low  fraction is due to the fact that many INTEGRAL AGN are newly discovered galaxies, often located on 
the Galactic Plane, i.e. in the zone of avoidance and therefore  their morphology is still poorly studied.
In the nearby Universe, elliptical galaxies are known to have less gas, dust, and ongoing star formation activity 
than spiral galaxies (\citet{Hirashita_2017} and refences therein).
Figure \ref{morph_nh} shows the histogram of each morphological class as a function of X-ray column density: while there seem to be no 
difference in absorption properties among different classes for column density below LogN$_{H}$ = 24, some difference is evident above,
since no Compton thick AGN is hosted in an elliptical host galaxy. 
However, this may be due to low number statistics, given the limited number of ellipticals considered here as they count for only  14\% of the entire INTEGRAL AGN sample. 
The fraction of heavily absorbed AGN is around 11\% for both spiral and lenticular galaxies;
if the same fraction is applied to ellipticals, then we would expect 3-4 objects in this class to be Compton thick, while none is observed.\\
The evidence of  a lower  fraction  of  Compton thick  AGN  in hard X-ray radio galaxies, typically hosted in elliptical galaxies, 
has already been discussed by \citet{Panessa_2016} and \citet{Ursini_2017} and is definitely an issue that deserves  
further investigation. \citet{Ursini_2017} also noted that a significant role in the absorption of  heavily absorbed radio galaxies 
could be played by material different from the classical pc-scale torus such as that traced by 21 cm H$_{I}$ absorption, which can 
be located much farther away, e.g. at galactic scales. 
Hence observational evidence even questions the origin of the X-ray absorption  in ellipticals and hints at a possible 
difference between the average properties of ellipticals compared to lenticular/spiral galaxies, especially in terms of Compton thick absorption.

\section{Host galaxies inclination}
\begin{figure}
	\includegraphics[width=\columnwidth]{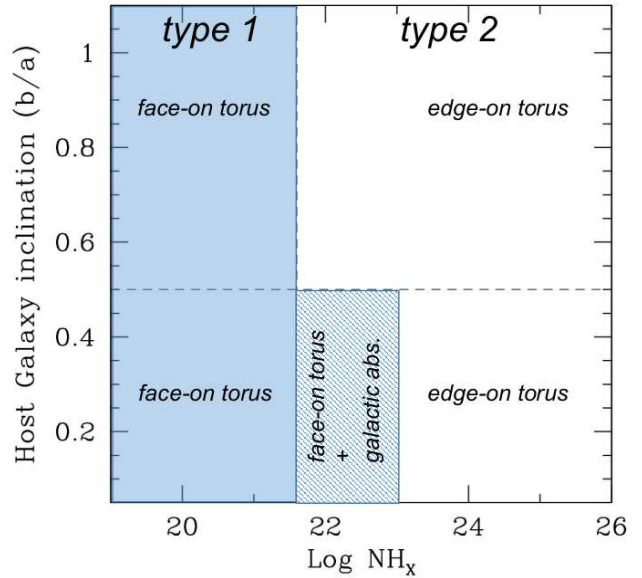}
  \caption{Scheme of the expected type of absorption in the b/a - N$_{H}$ plane for broad line (left blue side) and narrow line AGN (right side)
  in relation to the torus/host galaxy misalignment. The dashed blue region represents values where a passible type 1 AGN could be classified as type 2 due to
  absorption located in the host galaxy.  }
    \label{scheme}
\end{figure}

Using the large INTEGRAL AGN sample, we explore in this section the axial ratio distribution of broad  and narrow line AGN: 
figure \ref{histo1} displays the  histograms of $b/a$ by  galaxy type and within each type divided in morphological classes.
Note that the lack of galaxies with   $b/a <$ 0.2 is  due to the  non zero thickness of the disc which accounts for the fact 
that, even when seen edge-on, a galaxy has $b/a$ greater than zero \citep{Hubble_1926}.
Among the 122 type 1 AGN  of  our sample for which we obtained axial ratio information, only 12 ($<$10\%) have a host galaxy axial ratio below 
0.5; this percentage is much higher in type 2 AGN where 50 out of  178 type 2 objects (or 28\%) have  $b/a <$ 0.5{\bf\footnote{It is worth noting that the absorbed sources with no optical class considered as type 2, have axial ratio distributed over the whole range of b/a, and the only unclassified source included in the type 1 class, has b/a = 0.58.}}. 
By performing a KS test, we find that the axial ratio distributions of type 1 and type 2 AGN are different, 
since the probability that the two samples come from the same distribution is rejected at a confidence level of 99.99\%.
In order to quantify the deficit of type 1 AGN at low inclination angles, we compare the two histogram distributions.
Taking the difference between the two histograms, and assuming that above an inclination angle of 0.5 they are from identical distributions, we normalise it  so that the total difference in number of sources above that value is zero. In order to have a difference of zero below that threshold, 38 $\pm$ 11 sources must be missing from the type 1 distribution. This implies that in this sample there is a deficit of about 24\% ($\pm$ 5\%) of type 1 AGN, in agreement with \citet{Simcoe_1997} who found 30\% for an X-ray selected sample.\\
Furthermore, using the information on  the host galaxies morphology  and dividing the objects into early (Ellipticals and S0) and late  (Spirals) types, 
we also checked if the bias against edge-on host galaxies of broad line AGN is related to any specific  morphological type or 
is present in both types. 
Using (here and in the following) a confidence level of 99\% and applying the KS test, we find that 
the two distributions (type 1 versus type 2 AGN) are different both for early type as well as late type objects. This is also evident 
in figure \ref{histo1} where the different morphology types have been highlighted for type 1 and 2 AGN respectively. 
We therefore confirm that, even using AGN selected above 20 keV, the  deficit of type 1 AGN hosted in edge-on galaxies  exists in,
and is  independent of, all morphological types. \\
In addition to this deficit, it is interesting that the type 2 are distributed over the whole range of $b/a$.
This result confirms previous findings but also highlights the fact that this lack of Seyfert 1 in edge-on galaxies is not due to observational bias. 
As pointed out by \citet{Gelbord_2006}, if the torus was the only obscuration matter, then two possible scenarios should be taken into consideration: 
the first where the torus and the host galaxy plane are aligned and the second
where the two are misaligned. In the first case we should see a distribution of type 1 Seyferts peaked in face-on galaxies  and that of the type 2 Seyferts in edge-on galaxies 
while in the second case no correlation  between  Seyfert type and host galaxy inclination would be expected. 
From the studies performed so far and confirmed herein, neither of these two scenarios are compatible with the observational evidence since the broad line 
AGN have a strong correlation with the host inclination angle while the type 2 are broadly distributed over the whole $b/a$ range.
Our findings indicate that something intervenes  outside the nucleus of the AGN (on hundreds of pc scales), likely in the host galaxy; 
this absorbing material, which could be not aligned with the torus, can contribute to the final column density measured in the X-ray band.\\
From recent ALMA observations of some AGN, we have an indication that the molecular discs, or tori, detected at 10 pc scales
are kinematically decoupled from their host galaxy disc and have random orientations  \citep{2019A&A...623A..79C}.
Assuming a simplistic approach, i.e. not considering  the BLR but only the host galaxy and the torus, depending on their relative alignment or misalignment we expect the following 4 configurations.
If the torus is edge-on, the source is always classified as a type 2 AGN and should present X-ray absorption, whether or not the galaxy is seen  edge-on or face-on.
If the torus is seen face-on, then  we should expect a type 1 AGN configuration with no or very mild absorption unless the galaxy is seen edge-on and has absorbing 
material on large scale. 
In this last case the source could be  wrongly classified as a type 2 AGN although the absorption is not related to the torus.
This is schematically shown in figure \ref{scheme} where the horizontal line represents  the dividing line between face-on and  edge-on host galaxies at b/a = 0.5,
while the first vertical line represents instead the value of N$_{H}$ = 4 $\times$  10$^{21}$ cm$^{-2}$ which has been assumed 
as the dividing line between absorbed/unabsorbed objects. 
This value has been taken from \citet{Mateos_2016}  and corresponds to the extinction level capable of hiding the BLR. 
The dashed region delimited by log N$_{H}$ $\leq$ 23 and b/a $\leq$ 0.5, represents  the area where misclassified type 1 AGN could be located (i.e.
galaxy absorption in an edge-on host is sufficient to hide the broad line region).
With this scheme in mind, we can now investigate the relation between host galaxy inclination and X-ray absorption.

\subsection{Possible bias on redshift}
\begin{figure}
	\includegraphics[width=\columnwidth]{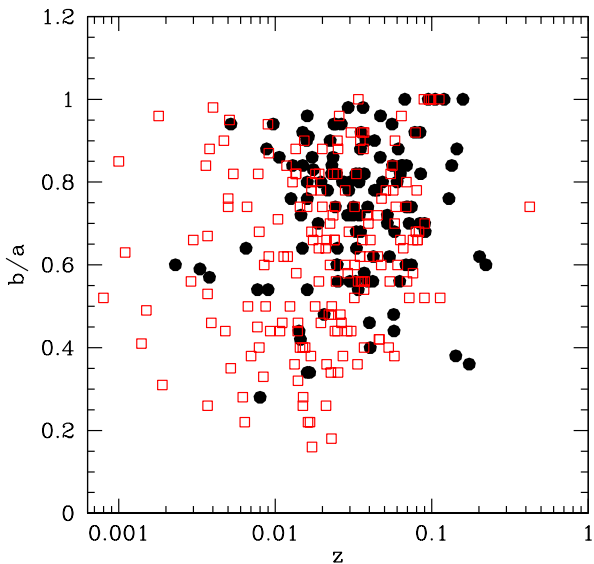}
  \caption{Redshift values of the INTEGRAL AGN plotted against the axial ratios of their host galaxies. 
           Filled black circles are type 1 AGN while open red squares are type 2.}
    \label{ax_zeta}
\end{figure}

\begin{figure}
	\includegraphics[width=\columnwidth]{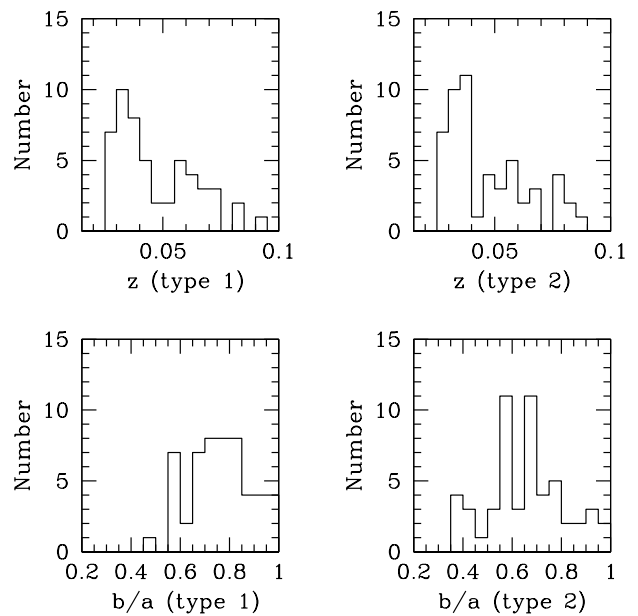}
  \caption{In the upper panels the distributions of redshift values in type 1 (left)  and type 2 (right) in the bin  0.026 and 0.095, while
  in the bottom panels the distributions of their relative axial ratios.}
    \label{histo_z}
\end{figure}

Before drawing any further  conclusions from the observed axial ratio distribution, any possible  bias introduced by the sample redshift distribution
must be checked.
It is important to investigate this issue since there are two considerations that pertain to the use of a non  complete sample of AGN. 
The first is that at high redshifts, the sources become point-like and therefore it is  very difficult to estimate the host galaxy 
axial ratio. The second concerns the widely known fact that in the local Universe type 2 objects outnumber type 1 sources by a factor 
of $\sim$4 and more \citep{2015MNRAS.454.1202S};  whether this  effect comes from  cosmological reasons or not is still  matter of debate.   
These two effects are highlighted in figure \ref{ax_zeta} where it is clear that the distributions of type 1 and type 2 galaxies are different 
in the $z$ -- $b/a$ plane: there are considerably more type 2 AGN  at low $z$ values than type 1 and  very few objects at high redshifts have 
axial ratio measurements.
In order to minimise these  effects, we have decided to analyse a subset of the two AGN populations restricting  both high and low $z$ values. 
To do this, we first choose all  sources between a lower and higher limit of redshift. The distribution of type 1 and type 2 objects between these 
two limits are then compared and the KS statistic applied to identify if they come  from the same population. 
This is done for all possible combinations of the two limits  in $z$. The combination which provides  the highest probability that the 
two populations come from the same  $z$ distribution (at 99.9\% confidence) are then further reduced to give the largest total number of objects. 
In this manner, we found that allowing the redshift to range from 0.026 to 0.095, we obtained a total number of 133 AGN with $z$ within this range;
of these AGN,  66 are of type 1 and 67 of type 2, with a probability of 99.97\% that they come from the same overall redshift distribution 
(top panels of figure \ref{histo_z}).
Next, the distribution of the axial ratio of these 133 sources divided in two optical classes were compared as shown in figure \ref{histo_z}  (lower panels).
Again the two distributions are found to be different,  with type 1 objects mainly located in face-on galaxies and type 2 spread over 
the entire range of   $b/a$ values. Using the KS statistic we find that the likelihood that the two populations come from the 
same distribution is rejected at a confidence level of 99.75\% implying that no bias in redshift affects our result.

 \begin{figure*}
 \centering
\includegraphics[width=12cm]{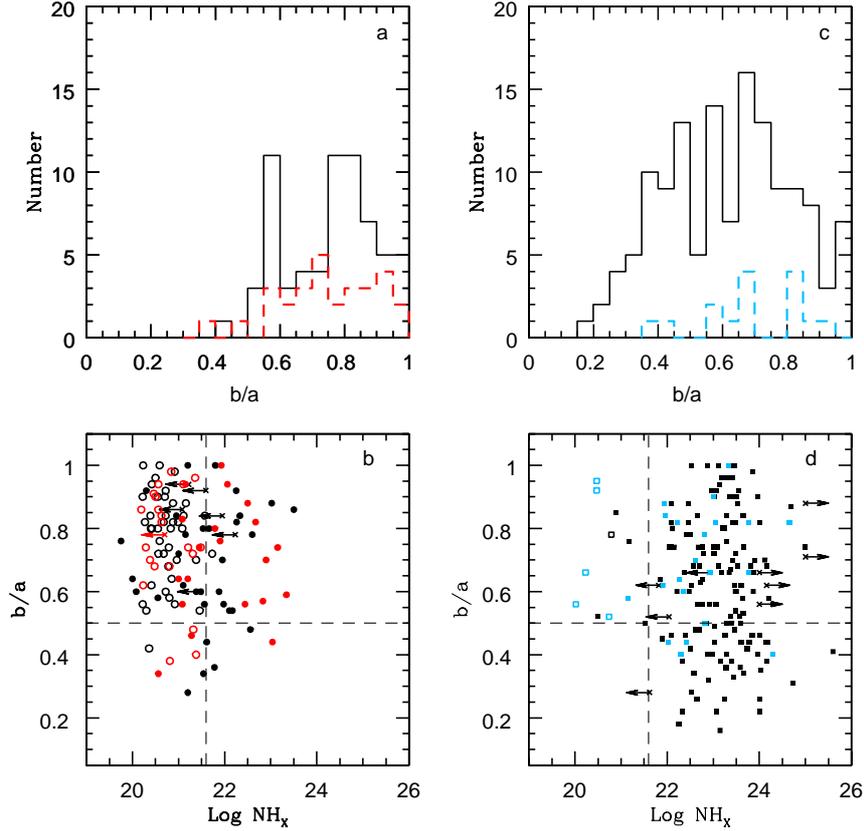}
   \caption{Upper panels: distribution of host galaxy axial ratios. Panel a) Seyfert 1/1.2 (black line) and intermediate Seyfert 1.5 (red dashed line); panel b): Seyfert 2 (black line) and intermediate Seyfert 1.8/1.9 (blue dashed line).
Lower panels:   axial ratios versus X-ray absorption: c) broad line AGN (black circles Seyfert 1/1.2 and red circles Seyfert 1.5; panel d) narrow line AGN, black squares Seyfert 2 and blue squares Seyfert 1.8/1.9. Open circles refer to N$_{H}$ upper limits or objects with no intrinsic absorption, i.e. only Galactic. Lines are like in figure 4.}
    \label{histo2}
\end{figure*}

\section{Host galaxy inclination  versus X-ray absorption} 
In order to investigate  the deficit of type 1 AGN in edge-on galaxies in more detail, we first considered for  each AGN type their optical sub-classes 
which is a way to investigate if and how the broad line region is detected.
In figure \ref{histo2} panel a), we compare  the axial ratio distribution of Seyfert  1-1.2 (solid black line ) and Seyfert 1.5 (dashed red line). 
As is evident in the figure but also quantified by the KS test, the two distributions are  similar (at a confidence level of 99\%) indicating 
that the deficit of broad line AGN in edge-on galaxies is present in  all  subclasses.\\
Equally, in panel b) of figure \ref{histo2}, the distribution of axial ratios in Seyfert 2 (black solid line) 
and Seyfert 1.8-1.9 (blue dashed line) are plotted in order to see if they are distributed differently and,  as is clear from the histograms and quantified by the 
KS statistics, they are not.
Also including all the sources of intermediate class (1.2, 1.5, 1.8 and 1.9) in the type 1 sample and leaving in the type 2 only the pure Seyfert 2 
objects, the two distributions resulted statistically different. The same result has been found considering only the high energy selected AGN belonging to
the INTEGRAL complete sample \citep{2009MNRAS.399..944M}.\\
In figure \ref{histo2} panel c) we plot  the axial ratios of type 1 AGN (Seyfert 1-1.2 black circles and Seyfert 1.5 red circles) versus 
the X-ray  absorption, the open circles indicate no intrinsic absorption (i.e. only Galactic) or upper limits.
Note that in  panels  c) and d) of figure \ref{histo2} the horizontal and vertical dashed lines follow the scheme introduced in figure \ref{scheme}.\\
As expected from the scheme of figure \ref{scheme}, type 1 AGN fall mostly in the left, top side of panel c), i.e. correspond to the situation 
where the torus and the galaxy are both seen face-on and no or mild X-ray absorption is present.
Regarding those type 1 AGN which, despite having torus and  galaxy both face-on, display X-ray absorption (i.e.
fall in the right upper side of panel c)) we assume that they can be largely explained as objects where the absorbing material 
is due to ionised gas located in accretion disc winds or in biconical structures close to the nucleus  \citep{Malizia_2012}.
A more complex interpretation pertain to those type 1 AGN which are hosted in edge-on galaxies and therefore fall
in the bottom parts of panel c).
First we notice that their number is limited  as expected on the basis of the consideration made in section 4.
These low numbers can either be due  to a low probability of having a configuration with a face-on torus in an edge-on 
galaxy or a high probability that  absorbing material in a host galaxy viewed edge-on is sufficient  to hide 
the BLR and to misclassify the object as a type 2 AGN.
Since there is no reason why each of these possibilities is likely, it is possible that  both  concur to create 
the observed low number of type 1 AGN seen  in edge-on galaxies.
In particular we already know of type 1 AGN with absorbing material hosted in edge-on galaxies, which is however not high 
enough to hide the BLR. These are the objects located in  the bottom left part of panel  c); one of these AGN is IC 4329A 
where the presence of a dust lane observed in the equatorial plane of the host galaxy
has been invoked to explain the source's mild X-ray neutral absorption which is however below our limit of  N$_{H}$ = 4 $\times$ 10$^{21}$ cm$^{-2}$
 \citep{Steenbrugge_2005}.
The most particular  objects are the 3 AGN, IGR J05347-6015, 4C +21.55 and ESO 140-G43,  in the bottom right corner of the panel c). 
Although they have a column density Log N$_{H} >$ 21.6 and are hosted in low inclination angle galaxies,  we still see their BLR.
However, these objects are either poorly studied as in the case of IGR J05347-6015 and 4C +21.55 
(only Swift-XRT short observations were available to estimate the X-ray column density) or found to be absorbed 
by complex and  ionised gas as for ESO 140-G43 (\citealt{Ricci_2010}).
In any case given the uncertainty  on the column  density of  these 3 objects, and given their proximity to the chosen $b/a$  
and LogN$_{H}$ boundaries, no firm conclusion can be drawn.\\ 
In  panel d) of figure \ref{histo2}, the X-ray column density is plotted against  the axial ratio of  Seyfert 2 (black squares) 
and Seyfert 1.8-1.9 (blue squares) and,  as for the type 1 AGN,  open symbols indicate no intrinsic absorption or upper limits.
In these objects the torus is by definition seen edge-on independently if the AGN is hosted in a face-on or an edge-on galaxy 
and their absorption is high: hence these objects  are expected to spread on the  right side of the diagram over all values of b/a.
Here the unusual objects are those  found in the upper left corner of panel d), which have a column density too  low 
to completely hide the BLR. Not considering  Sey1.8-1.9 types where an intrinsically variable ionising continuum or absorption/reddening unrelated 
to the torus can explain their intermediate classification, there are 4 objects: NGC 4736, IGR J19260+4136, IGR J03249+4041-SW and
IGR J14515-5542, optically classified as Seyfert 2, which have no absorption in X-rays.
NGC 4736 is a nearby star-forming ring galaxy \citep{van_der_Laan_2015} where the torus is probably not present due to the low 
luminosity of the nucleus; star formation could then be responsible for the type 2 optical class. 
IGR J03249+4041-SW is one component of a pair of Seyfert 2 in close interaction (pair  distance around 12 Kpc) and it is therefore possible 
that its type 2 class  is  simply due to dust associated to the galaxy merging.
The other two AGN are poorly studied, but their column density is well measured and confirmed using XMM-INTEGRAL data for  IGR J14515-5542 \citep{De_Rosa_2012}
and  NuSTAR data for IGR J19260+4136 (our analysis) respectively. 
The fact that no prominent iron line is seen at soft X-ray energies excludes that these 
are Compton thick AGN (see \citealt{Malizia_2007}); it is  therefore very likely that they belong to the small fraction of 
naked Seyfert 2 galaxies where the torus is missing and the optical classification is due to other circumstances \citep{Panessa_2002}.
Otherwise they can be type 1 AGN where the combined effect of a very broadened emission profile and an intrinsic weakness with respect to the host galaxy
conspire to produce  a type 2 spectrum \citep{2019MNRAS.488L...1B}. \\
The objects in the bottom right corner of panel d) are  the most relevant for the aim of this work. Following the scheme of
figure \ref{scheme}, for these type 2 objects we cannot discriminate between absorption due to  material located on large scales and related to the host galaxy,
and that due to the torus.

\section{Discussion}
It is a well known effect that optically selected type 1 Seyfert tend to avoid edge-on host galaxies.
Extensive studies have been performed in different wavebands in order to explain this bias \citep{Keel_1980,Simcoe_1997,Lagos_2011} and although some 
objects missed in UV and optical surveys have been recovered going at higher energies, the deficit of type 1 hosted in edge-on galaxies still remains. A definitive test is that provided by the use of a hard X-ray selected sample which is less biased against obscured objects and therefore free of the limitation which affects surveys at other frequencies (i.e. from optical to soft X-rays). The first test at high energies was performed by \citet{winter_2009} on a limited Swift/BAT sample of around 80 AGN and also with this selection, they found that edge-on galaxies only host the more absorbed AGN, finding that from the 11 edge-on sources in their sample, only one is associated with a Sey 1.\\
In this work using the INTEGRAL high energy selected sample of AGN (376 Seyferts), we have confirmed that there is a deficit  of around 24\% ($\pm$ 5\%) of type 1 objects hosted in edge-on galaxies.
Since this bias is not observational, i.e. due to the particular selection of the AGN sample, we deem that this effect 
is most likely physical and presumably due to the presence of absorption located in the large scale structure of the host galaxy.
This extra absorption that we called for simplicity galactic, contributes to the total column density observed in X-rays and, even alone, could be sufficient to hide 
the broad line region providing a type 2 classification. 
The observed axial ratio distribution found in the present analysis can only  be  explained by considering a galactic absorption
lying on the galaxy plane (e.g. inner bars, rings, dust lanes) in addition  to  the nuclear absorber.\\
In this picture, the role played by the orientation  of the two absorbing structures (large and small scale) is obviously more complicated than 
in the more conventional unified picture where only the orientation of the torus determines if a source  is classified as a type 1 or 2.
Within this scenario either absorber is capable of attenuating the BLR flux, and so not only the orientation of each with respect to 
the line of sight is important but also their relative location and orientation.
Following this scenario, when the torus and the galactic absorber 
are both edge-on, much of the absorbing material lies in the shadow of the torus leaving a wide range of angles which provide a direct view 
of the nucleus and therefore a type 1 classification.
On the contrary, if the two absorbers are severely  misaligned,  then the absorbing 
material, either due to the torus or the galactic absorber, covers a much larger fraction of the nucleus, substantially increasing the 
probability that a randomly oriented observer will see a type 2 AGN. 
In this case we can have 3 configurations that provide a type 2 classification:
one in which only the torus  is detected, one  where both absorbing structures lie along the line of sight and, finally, one where only a thick layer of galactic absorber is intercepted.
In the first two cases the classification is that of a classical (torus hidden) type 2 AGN where the galactic absorption can eventually 
enhance the X-ray column density due to the torus.
In the last case the source is formally not a type 2 AGN according  to the classical unified theory, since the torus is not along the line of sight, but 
if  LogN$_{H}>$21.6 and the host galaxy is edge-on, the BLR becomes hidden and the source is optically classified as a type 2 AGN.
In other words, if it were  not for the host galaxy, these would be classified as type 1, i.e. it is the host galaxy that makes them appear as type 2.
Given their relevance, it is important to find some examples of these AGN within our large data-set.
If we assumed that the parsec scale tori are  always Compton thick or nearly so,  i.e. have column density above a few 
10$^{23}$ cm$^{-2}$,  then objects with absorption much lower are probably hidden behind galactic material.
Therefore in order to search for {\it "misclassified"} type 1 Seyferts, we  focus on those 
Seyfert 1.8, 1.9 and 2 hosted in edge-on galaxies (i.e. having  $b/a$ below 0.5) and have mild X-ray absorption 
(below 10$^{23}$ at cm$^{-2}$), i.e. those located in the dashed region of figure \ref{scheme}.
There are 22 such objects in our sample of INTEGRAL AGN, most of which are newly discovered and therefore poorly studied.
For 9 of them broad band information could be gathered and analysed to verify their classification. 
These objects are shown in Table 1 where we list their name, presence of bar or dust lanes as  reported in 
the literature and evidence, from the near infrared spectroscopy (NIR), for the presence of broad emission lines. 
The NIR regime is particularly useful from this purpose since, being less affected by dust extinction, is more likely  to 
show a BLR otherwise hidden in the optical band.
The  sources listed in Table 1 are clear examples of how a type 1 nucleus can hide behind absorbing structures 
that are located in the host galaxy.
Some objects  like NGC 7172 \citep{Smaji__2012}, NGC 2992 \citep{Trippe_2008}, NGC 5252 \citep{Kotilainen_1995}  and
NGC 5506 \citep{Nagar_2002} have already been discussed in the literature as examples of how a Seyfert 1 can be 
disguised as a Seyfert 2 due to galactic obscuration, while the other examples are  suggested here for the first time. 
Although with the available observations their number is small (these objects represent 40\% of the sample) and not yet sufficient  to compensate
for the lack of type 1 AGN in edge-on galaxies, their existence indicates this kind of {\it "misclassification"} due to galactic absorption can take place and, if extended to a large 
number of objects, might explain  the difference in the  axial ratio distributions of type 1 and 2 AGN.
On the other hand the existence of such objects gives support to the idea 
that an absorption located at larger scales is at work within a large fraction of local AGN.\\
As said, now that the co-evolution of galaxies and black holes is well established (e.g. see \citet{2014ARA&A..52..589H}  and references therein) 
and the advent of the ALMA observations  probe and image the gas within 100 pc of the AGN, there is beginning to accumulate evidence of multi absorption components.
Mass and gas concentrations in host galaxies which are responsible for the 
feeding mechanism in AGN, can also contribute to the obscuration of  their nuclei.
The mechanism appears to be that of kinematically decoupled embedded bars, i.e. the combination of a slowly rotating kpc-scale 
stellar bar and a kinematically decoupled nuclear bar, with overlapping dynamical resonances \citep{Combes_2013}. 
Such resonances and kinematic decoupling are fostered by a large central mass concentration and high gas fraction. 
The gas is first halted in a nuclear ring (a few 100 pc scale), and then driven inward under the influence 
of the decoupled nuclear bar obscuring the central engine. 
This has been observed in ESO 428-G14, where the measured kinematics is consistent with a nuclear inflow, or inner bar, 
which feeds the AGN \citep{2020ApJ...890...29F}.\\
Therefore the picture emerging from recent observations is that this galactic absorption may reside in different structures present in the host galaxy 
which include dust lanes, bars, rings etc. In this paper  we focus on their contribution to the  obscuration of the direct radiation from the AGN 
especially when the host galaxy is seen edge-on.
With this in mind also Compton thick AGN could be explained as sources in which the two 
absorbers on galactic and nuclear scales  are both aligned and seen edge on.
A similar result was also obtained by \citet{Goulding_2012}, studying the deep silicate absorption features seen in many  
Compton-thick AGN. \citet{Goulding_2012} found that an important  contribution to the observed mid-IR  extinction in these objects is dust located 
in the host galaxy, i.e. due to disturbed morphologies, dust lanes or galaxy inclination angles.
However, these authors  considered these  two absorbers, galactic versus torus, as two alternative absorbing structures while 
here we assume that they could be both present in the same object and possibly aligned to enhance the nuclear obscuration. 
To produce Compton thick absorption our line of sight intercepts an edge-on torus and the  galactic absorber (like for example an inner bar) can intervene or not 
depending if it is  present or not in the host galaxy.
The other possible geometry, i.e. that of a face on torus, either  aligned or not with the galactic absorber,  will always produce a Compton thin source \citep{2019arXiv190713137B}.
Therefore we can say that a Compton thick AGN can have two flavours: one in which our line of sight intercepts 
only the torus and one in which both the torus and the galactic structure are seen edge-on. 
This last configuration produces objects like ESO 428-G14 and those  discussed by \citet{Goulding_2012}.  
We note  that the deep silicate absorption features observed in \citet{Goulding_2012} sample, cannot be explained 
in terms of torus absorption models (see for example \citet{Garc_a_Gonz_lez_2017} ) and can only be produced by dust on 
galactic scales.

\section{Summary and Conclusions}
Using the hard X-ray selected sample of AGN detected by INTEGRAL/IBIS, in this work we have investigated  the possible 
contribution of  absorbing material located on a scale of hundreds of parsecs, i.e. in the host galaxies, to the total amount of 
N$_{H}$ measured in the X-ray band.
By collecting  the axial ratios  ($b/a$) of the host galaxies  for all our sample sources,  
we have verified that also within our hard X-ray selection sample there is a deficiency of around 24\% ($\pm$ 5\%) of type 1 AGN hosted in edge-on 
galaxies (those with $b/a < $0.5).
We have investigated the distribution of the host galaxies axial ratios in type 1 and type 2 AGN and further highlighted 
the optical Seyfert subclasses (type 1.5 for the unabsorbed and type 1.8 and 1.9 for absorbed AGN) to check whether there 
is a trend of the optical class with the inclination of the galaxy, in other words, we checked for the presence of 
absorption able to hide the BLR, or part of it, in the different galaxy axial ratios, and we found none.
Possible bias in redshift has been excluded. We found the same effect in a well determined range of 
$z$ (0.026 $< z <$ 0.095) where the number and the distributions of the two classes are statistically 
the same (at 99.9\% confidence).
Clearly this indicates that some material located in the host galaxy on scales of hundreds of parsec and not aligned 
with the putative absorbing torus of the AGN can contribute to the column density measured in the X-ray band.
In particular we have developed a scheme of the expected AGN type as a function of X-ray absorption and axial ratio
in the different configuration of torus and galaxy inclinations: when the torus is seen edge-on, we always have a type 2 absorbed 
AGN independently of the host galaxy orientation; when the torus is seen face-on we may have a  type 1 AGN unabsorbed or mildly 
absorbed or  "misclassified type 2 objects where the absorption is on galactic scales.
Plotting the axial ratio versus the column density for type 1 and type 2 we have highlighted peculiar sources and  in particular have been able to 
identify a set of possible "misclassified type 2 AGN", i.e those which are absorbed (type 2) and located in edge-on
galaxies.
Our conclusion is that these absorptions, galactic versus torus, are not alternative but  could be both present in the same object
and possibly aligned to enhance the nuclear obscuration. 
Within this scenario a Compton thick AGN can come in  two flavours: one in
which our line of sight intercepts only the torus and one in which
both the torus and the galactic structure are seen edge-on.

\begin{table*}
\begin{center}
\caption{Possible Misclassified type 1.9/2 AGN}
\begin{tabular}{lcccc}
\hline
{\bf name}                                      &  {\bf dust$^{\dagger}$/bar}  &   {\bf ref} & {\bf NIR BLR } &  {\bf ref}   \\
\hline                 
IGR J02343+3229= NGC 973       & DC/N               &      1       &      --          &                 \\
IGR J09025-6814=UGC11397      &  --/Y                 &      --       &      --          &                \\
NGC 2992                                     & DC/N               &       2       &     Y           &        3       \\
NGC 5506                                     & D-S/Y              &       4       &     Y            &       5       \\
NGC 5252                                     & R-I/N               &        2       &     Y           &        5       \\
IGR J19039+3344=NGC2788A    &  D/Y                 &        6       &    -              &                \\
NGC 7172                                     & DC/N               &       2        &    Y             &       7       \\
NGC 7314                                     & DC/Y                &       2        &   Y             &       5, 3   \\
IGR J22367-1231=MKN 915        & DC/N                &       2        &    Y            &       5       \\
\hline
\end{tabular}
\end{center}
($\dagger$) Same nomenclature as used by \citet{Malkan_1998}.
Ref: (1) \citealt{Verstappen_2013}; (2) \citealt{Malkan_1998}; (3) \citealt{Onori_2016};  (4) \citealt{Martini_2003}; (5) \citealt{Lamperti_2017}; (6) NED; (7) \citealt{Smaji__2012}.
\label{tab1}
\end{table*}

\section*{Acknowledgements}
The authors acknowledge financial support from ASI under contract n.  2019-35-HH.0.
We acknowledge the usage of the HyperLeda database (http://leda.univ-lyon1.fr).
This research has made use of the NASA/IPAC Extragalactic Database (NED) and NASA/ IPAC Infrared Science Archive, which are operated by 
the Jet Propulsion Laboratory, California Institute of Technology, 
under contract with the National Aeronautics and Space Administration.
Acknowledgements go to Rick White at Pan-STARRS database for his help in finding/interpreting  PS1 tables.
We thank Chiara Feruglio for worthwhile discussion. The authors  also thank Marco Mignoli for useful discussion on 
AGN/host galaxy characteristics in optical/IR band, Eliana Palazzi and Andrea Rossi for imaging analysis in optical/IR band.\\
We thank the anonymous referee for useful remarks which helped us to improve the quality of this paper.




\bibliographystyle{aa}
\bibliography{angela_host}



\appendix

\section{New INTEGRAL AGN }
Here we discuss the new AGN  reported by \citet{Mereminskiy_2016} in deep extragalactic surveys and those reported 
by \citet{Krivonos_2017} from an extensive galactic plane mapping.
Among the 147 sources reported by \citet{Mereminskiy_2016}, 70  are new hard X-ray emitting objects never  reported in 
previous INTEGRAL surveys (\citealt{Bird_2016} and references therein): 23 of these are still unidentified and therefore have 
not been considered.
Of the  remaining 47 objects, all optically classified as AGN, 8 are blazars and 7 have no X-ray observation available 
to characterised their 2-10 keV spectra; they have therefore been excluded from the list of AGN considered in this work. 
The  remaining 33 objects have instead been added to the large database of INTEGRAL AGN   (\citealt{Malizia_2012, Malizia_2016}). 
Regarding instead the new hard X-ray sources reported  along  the Galactic Plane Survey  by  \citet{Krivonos_2017}, 
21 are identified with AGN by the authors: 11 are unambiguously classified as Seyfert galaxies and therefore have been 
added to the INTGRAL AGN sample, while the rest, being  either blazars or objects of unknown class, have  been omitted. 
All these new entries (33 + 11) have been listed in table \ref{mere}.2 and \ref{krivo}.3 together with their coordinates, redshift, optical class 
and X-ray spectral parameters (column density, photon index, 2-10 and 20-100 keV fluxes) for uniformity with the work 
of \citet{Malizia_2012, Malizia_2016} . 
It is worth noting that also these new addition have been unambiguously associated with their X-ray counterparts thus allowing 
to restrict their positional error box and therefore to be optically identified and classified with good confidence. 
To provide the X-ray spectral parameters  for these new 44 AGN,  we have checked both the literature and the archives to search 
for information.
In many cases, the X-ray data analysis has already been performed and the results reported in various publications as listed 
in the last column of Table \ref{mere}.2 and \ref{krivo}.3.
For 7 objects the X-ray data analysis is performed and presented here using archival Swift/XRT observations; in a couple 
of cases we preferred to re-analyse  the data since more exposure or not convincing results were found 
(e.g.  IGR J21099+3533 and IGR J21382+3204 already analysed by \citealt{Ricci_2017}).
The 2-10 keV spectral analysis has  been performed following the method described in \citet{Malizia_2016}; the 20-100 keV 
fluxes have been converted from  the 17-60 keV flux reported in \citet{Mereminskiy_2016} and \citet{Krivonos_2017} by assuming 
a simple power law $\Gamma$=2.01$\pm$0.04, as found by \citet{Molina_2013}.

\clearpage

\begin{center}

\begin{table*}
\caption[]{INTEGRAL/IBIS  AGN}
 \begin{tabular}{|| l c l l r l | c }
  \hline\hline
{\bf Name  }                           & {\bf $z$ }  &  {\bf Class} & Log N$_{H}$ & {\bf b/a}    &  Morph/BAR \\
\hline
 IGR J00040+7020                &   0.0960     &    Sy2          & 22.52             &   1.00     &      --     \\
 IGR J00256+6821                &   0.0120     &    Sy2          & 23.60             &   0.62     &      --       \\
 IGR J00333+6122                &   0.1050     &    Sy1.5       & 21.93             &   1.00     &      --                               \\
 SWIFT J0034.5-7904           &   0.0740     &    Sy1          & {\bf 20.79}      &    0.74    &      --          \\
 IGR J00465-4005                 &   0.2010     &    Sy2          &  23.38            &   --          &    --      \\
 Mrk 348                                 &   0.0153     &    Sy2         &  23.02            &  0.90      &   SA     \\
 Mrk 352                                 &   0.0150     &    Sy1         & {\bf 20.72}      & 0.92       &  S0A     \\
 Mrk 1152                               &   0.0537     &    Sy1.5      & {\bf 20.23}      &  0.62      &  S       \\
 Firall 9                                   &   0.0470     &    Sy1.2      & {\bf 20.50}      &   0.96     &  S0       \\  
 NGC 526A                            &   0.0191     &    Sy1.9      &  22.23             &   0.82     &  S0A    \\
 ESO 297-18                          &   0.0252     &    Sy2         &  23.66             &   0.34    &  S        \\
 IGR J01528-0845                 &   0.0370     &    Sy2         & 23.49              &  0.54     &  SB      \\      
 IGR J01528-0326                 &   0.0172     &    Sy2         & 23.15              &  0.16     &  SA            \\
 IGR J01545+6437                &   0.0349     &    Sy2         &  $<$21.82       &  0.62      &  --      \\
 Mrk 584                                 &   0.0788    &    Sy1.8      & {\bf 20.47}       &  0.92     &  --           \\ 
 NGC 788                               &   0.0136   &    Sy2          &  23.48             &  0.82      & S0A      \\
 Mrk 1018                               &   0.0424    &    Sy1         &  {\bf 20.41}      &  0.62     & S0              \\ 
 IGR J02086-1742                 &   0.1290    &    Sy1.2      & $<$20.33         &  --         &  --                \\ 
 IGR J02097+5222                &   0.0492    &    Sy1         & {\bf 21.23}       &   --         & --                 \\ 
 Mrk 590                                &   0.0264     &    Sy1        & {\bf 20.42}       &  0.94     & SA                \\ 
 SWIFT J0216.3+5128          &   0.4220     &    Sy2?      & 22.10              &  0.74     &  --            \\ 
 Mrk 1040                               &   0.0166    &    Sy1.5      & 20.56             &  0.34     & SA            \\ 
 IGR J02343+3229                &   0.0162    &    Sy2/LINER   & 22.34       &  0.22       & SA            \\ 
 NGC 985                              &   0.0431    &    Sy1.5         & {\bf 20.50}  &  0.90      &  SB?            \\ 
 NGC 1052                            &   0.0050    &    Sy2/LINER & 23.30         &  0.76       & E             \\ 
 RBS 345                               &   0.0690    &    Sy1            &{\bf 20.72}   &   0.84      &  --        \\ 
 NGC 1068                            &   0.0038    &    Sy2            & $>$25.00     &  0.88    & SB        \\ 
 QSO B0241+62                   &   0.0445     &    Sy1.2          & 21.32         &  --          & --              \\ 
 MCG -07-06-018                  &   0.0696    &    XBONG     & $>$24.00    &  0.56      &  S0A      \\ 
 SWIFT J0249.1+2627          &   0.0580    &  Sy2              & 23.43         &  0.70      &  --        \\
 IGR J02504+5443                &   0.0151     &    Sy2           & 24.20        & 0.28        & S0         \\ 
 MCG -02-08-014                  &  0.0167     &    Sy2?          & 23.08        & 0.22       & S       \\ 
 NGC 1142                            &   0.0288    &    Sy2             &  23.80       &  0.44      & E        \\ 
 MCG -02-08-038                  &   0.0326    &    Sy1             &  21.56       &  0.56      & SAB     \\
 NGC 1194                            &   0.0136    &    Sy2             &  24.20       &  0.58      & S0A  \\ 
 PKS 0312-770                      &   0.2252    &    Sy1/QSO   &  {\bf 20.93} &  --          &  --           \\
 B3 B0309+411B                  &   0.1340     &    Sy1              & {\bf 21.11}  &   0.84   &  --            \\    
 SWIFT J0318.7+6828         &   0.0901    &    Sy1.9           & 22.61        &  0.70       & --         \\ 
 NGC 1275                           &   0.0175    &    Sy1.5/LINER & 21.08      & 0.83        &  S0        \\ 
 1H 0323+342                      &   0.0610     &    NLS1          & {\bf 21.16}  &    0.88    &  --            \\ 
 IGR J03249+4041-SW       &  0.0477     &    Sy2              &  21.18      & 0.76           & --      \\ 
 IGR J03249+4041-NE        &  0.0475     &    Sy2             &  22.48       & 0.60           & S          \\ 
 IGR J03334+3718               &  0.0558    &    Sy1.5          & 21.15        &  0.94          & --         \\ 
 NGC 1365                           &   0.0054    &    Sy1.9          & 24.65       &  0.82          &  SB   \\ 
 ESO 548-G81                     &   0.0145    &    Sy1             &  {\bf 20.36}  & 0.42        &  SB         \\ 
 SWIFT J0353.7+3711         &  0.0186     & Sy2/LINER   & 22.57        & 0.68            &  S0      \\
 4C +62.08                            &  1.1090     &  Sy1              & {\bf 21.51}   &  --            &  --        \\
 SWIFT J0357.6+4153         &  0.0530     &  Sy1.9           &  22.30        &  0.40           &  --        \\
 3C 098                                &   0.0304    &    Sy2             &  23.08        &   0.92         &  E     \\ 
 4C 03.8                               &   0.0890    &    Sy2             &  23.45        &  1.00          &  E      \\ 
 3C 111                                &   0.0485    &    Sy1             &  21.66        &  0.80          &  E       \\ 
 IGR J04221+4856              &  0.1140     &   Sy1              & {\bf 21.85}   &     --          &  --      \\ 
 LEDA 15023                       &   0.0450    &   Sy2              &   23.48        &   0.72       &   E     \\ 
 3C120                                 &   0.0330    &    Sy1.5          &   21.20        &   0.64       & S0       \\ 
 UGC 3142                          &   0.0216   &    Sy1              &   22.60        &   0.78       & S0B      \\ 
 SWIFT J0444.1+2813        &   0.0113    &    Sy2              & 22.53          &  0.62           & S    \\
 SWIFT J0450.7-5813         &   0.0907   &    Sy1.5           &  {\bf 20.80}  &   0.68       &  --    \\
 MCG -01-13-025                &   0.0159    &   Sy1.2           &   {\bf 20.55}   &  0.90      & SAB       \\ 
 LEDA 168563                     &   0.0290    &    Sy1             & {\bf 21.73}      &  0.72    & --         \\ 
 SWIFT J0453.4+0404        &  0.0296    &    Sy2               & 24.16         &  0.68       &   E/S0A       \\            
 ESO 033-G02                    &   0.0181   &    Sy2               &   22.10        &  0.82       & S0B        \\ 
 LEDA 075258                     &   0.0160  &     Sy1              &   19.75        & 0.76        & E         \\  
\hline
\end{tabular}
\label{all}
\end{table*}

\begin{table*}
\begin{tabular}{|| l c l l r l | c }
  \hline\hline
{\bf Name  }                     & {\bf $z$ }  &    {\bf Class}   & Log N$_{H}$ & b/a   & Morph/BAT      \\
\hline
 SWIFT J0505.8-2348         &  0.0350   &    Sy2               &  23.50          &   0.90  &  --         \\ 
 IGR J05081+1722              &  0.0175   &  Sy2                 &  22.38          &   0.66   & S0        \\
 4U 0517+17                        &  0.0179  &   Sy1.5             &  20.95          &  --         & E/S0            \\ 
 SWIFT J0515.3+1854        &  0.0235  &   Sy2                 &  23.06          &  0.82     & E           \\
 Ark 120                               &  0.0327  &   Sy1                 &  {\bf 20.99}  &  0.82    & SA           \\
 SWIFT J0516.3+1928        &  0.0211  &  Sy2                  & 22.64           & 0.26     & S       \\ 
 SWIFT J0519.5-3140        &   0.0126  &  Sy1.5              &  21.90          & 0.76     & S0B            \\ 
 PICTOR A                         &  0.0351   &  Sy1/LINER     &  {\bf 20.78}    & 0.68     & S0A      \\ 
 PKS 0521-36                     & 0.0565    &  Sy1                &  {\bf 20.55}    & --          & S0             \\
 SWIFT J0544.4+5909      &  0.0659   &  Sy1.9              &  22.26             &  0.64   & --           \\      
 IGR J05470+5034            &  0.0360   &  Sy2                 & 23.18             & 0.66     & S0         \\
 NGC 2110                        &   0.0078  &  Sy2                  & 22.46             & 0.82     & S0AB      \\ 
 MCG+08-11-011              &  0.0205  &   Sy1.5               & {\bf 21.32}     &  0.48     & SB       \\  
 4U 0557-385                   &   0.0339   &    Sy1.2             &  22.10           & 0.54      &  S0           \\ 
 IRAS 05589+2828          &   0.0330     &    Sy1              & {\bf 21.66}     &  --          &  --          \\
 SWIFT J0601.9-8636     &   0.0064   &    Sy2                &  24.00           & 0.22        & S    \\ 
 IGR J06058-2755           & 0.0900     &    Sy1.5            &  {\bf 20.38}    &  0.70       &  --         \\ 
 Mrk 3                               &   0.0135   &    Sy2              &  24.00            &  0.82       & SAB?      \\ 
 IGR J06233-6436           & 0.1289    &    Sy1               & {\bf 20.59}     &   0.76       &  --       \\ 
 IGR J06239-6052           & 0.0405    &    Sy2               & 23.35            &  0.66        & S0B        \\ 
 SWIFT J0623.8-3215     &  0.0224   &    Sy2               &  23.91           & 0.70         & S0B           \\
 SWIFT J0640.4-2554     &  0.0248   &  Sy1.2             & 21.38             & 0.60         & S          \\ 
 IGR J06415+3251          & 0.0172   &  Sy2                 &  23.20            &  0.80        &  E          \\ 
 Mrk 6                               & 0.0188  &  Sy1.5               & 22.90            &   0.70        & S0AB          \\ 
 Mrk 79                             & 0.2219    & Sy1.2              & {\bf 20.72}     & 0.60         &  SB        \\
 IGR J07565-4139           & 0.0210    & Sy2                & 21.86              &  0.64        & --          \\ 
 IGR J07597-3842           & 0.0400   & Sy1.2             &  {\bf 21.78}       &   --           & --          \\ 
 ESO 209-12                    & 0.0405   & Sy1.5             &   {\bf 21.38}     &  0.40        & S       \\ 
 Mrk 1210                         &  0.0135  & Sy 2               &  23.52              &  0.88       & S         \\
 PG 0804+761                  &  0.1000   & Sy1               &  20.70              &    --        & E         \\ 
 IGR J08190-3835          & 0.0090     & Sy2               & 23.13               &   0.94      & --        \\ 
 FRL 1146                      & 0.0316    & Sy1.5              & 21.45               & 0.74        & S        \\ 
 SWIFT J0845.0-3531    & 0.1370    &   Sy1.2           & 22.38               & --             &  --       \\
 IGR J08557+6420         & 0.0370     &   Sy2?          & 23.29                &  0.40      & S?         \\  
 IGR J08558+0814         & 0.2200     & Sy1              &  {\bf 20.67}         &  --         & --      \\ 
 Mrk 18                            & 0.0111     &  Sy2              &  23.26               & 0.46      & S0     \\                                                         
 IGR J09025-6814          &  0.0140    &  XBONG       & 22.90               & 0.32        & SAB      \\ 
 IGR J09026-4812          &  0.0391    &    Sy1            &  21.98              &  0.56      &  --         \\
 1RXS J090431.1-382920 &  0.0160  &  Sy1             &  {\bf 21.46}       &  0.54      & S         \\
 SWIFT J0917.2-6221   &  0.0573    &  Sy1               &  21.61               &  0.44      & --       \\
 IGR J09189-4418          &    --         &   AGN             &  22.66               &  0.84      & --          \\
 MCG-01-24-012            &  0.0196   &    Sy2              & 22.80               &   0.78     & SAB        \\
 Mrk 110                         &  0.0353   &    NLS1           &  20.30              & 0.92       &  E         \\
 IGR J09253+6929        &  0.0390   &    Sy1.5           &  23.15              & 0.74       &  --        \\
 SWIFT J0929.7+6232   &  0.0256   &   Sy2               & 23.31              & 0.96        & E      \\
 IGR J09446-2636         &  0.1425    &   Sy1.5           &  {\bf 20.81}     & 0.38         &   --      \\
 NGC 2992                   &  0.0077   &    Sy2                & 21.90              & 0.45       & SA      \\
 MCG-05-23-016          &  0.0085   &    Sy2                & 22.21              & 0.60       & S0      \\
 4C 73.08                      &  0.0581   &   Sy2                 & 23.96              & 0.90       & --        \\
 IGR J09523-6231       &  0.2520    &  Sy1.9               & 22.80             & --             & --        \\
 M81                              & 0.0008     &  Sy1.8/LINER  &  {\bf 20.74}    &  0.52        & SA     \\
 NGC 3081                   &  0.0079    &    Sy2                & 23.82            & 0.68         & SAB       \\
 SWIFT J0959.7-3112  &  0.0370    &  Sy1                  &  {\bf 20.80}   &  0.58        & S0B           \\
 NGC 3079                    &  0.0037    &  Sy2                 & 24.00            &  0.26        & SB     \\ 
 SWIFT J1009.3-4250  &  0.0330    &    Sy2                & 23.41            &  0.82        & SAB     \\
 IGR J10147-6354        &  0.2020    &    Sy1.2             & 22.30            &   --           & --           \\
 NGC 3227                    &  0.0038   &    Sy1.5             &  22.83           &  0.57       &  SAB       \\
 NGC 3281                   &  0.0107    &    Sy2                &  24.18          &   0.44       & SAB       \\
 SWIFT J1038.8-4942  &  0.0600    &    Sy1.5             &  21.79         &   0.80        & --          \\
 IGR J10404-4625       &  0.0240    &    Sy2                 &  22.61         &  0.66        & S0       \\
 MCG+04-26-006         &  0.0200    &    LINER            & 23.09          &  0.74        &  SA           \\
\hline
\end{tabular}
\end{table*}

\begin{table*}
\begin{tabular}{|| l c l l r l | c }
  \hline\hline
{\bf Name  }    & {\bf $z$ }  &    {\bf Class}   & Log N$_{H}$ & {\bf b/a}  & Morph/BAT  \\
\hline
 NGC 3516                   &  0.0088   &    Sy1.5               & 22.50     & 0.88         & S0B           \\
 IGR J11366-6002        &  0.0140    &  Sy2/LINER      & 22.40      & 0.46         & --         \\
 NGC 3783                   &  0.0097   &    Sy1.5              &  22.06     & 0.94         &  SB        \\
 SWIFT J1143.7+7942  & 0.0065   &   Sy1.2               &  20.00     & 0.64         &  S0B        \\
 H1143-182                   &  0.0329   &    Sy1.5             &{\bf 20.48}  &  0.68     &    SA        \\
 PKS 1143-696              &  0.2440  &   Sy1.2              & {\bf 21.21}  &    --       &   --       \\
 B2 1144+35B               &  0.0631   &  Sy1.9               & {\bf 20.23}  &  0.66    & --        \\
 SWIFT J1200.8+0650  &  0.0360   &  Sy2                   & 22.92        & 0.56     & S0       \\
 IGR J12026-5349         &  0.0280   &    Sy2                 &  22.41       &  0.78    & S0B      \\
 NGC 4051                    &  0.0023   &    NLS1               & $<$21.48 &   0.60    & SAB    \\
 NGC 4074                    &  0.0224   &    Sy2                  & 23.48       & 0.60      & S0    \\
 Mrk 198                         &  0.0242   &   Sy2                  & 23.00        & 0.74      & S0AB     \\
 NGC 4138                    &  0.0030   &    Sy1.9               & 22.95        & 0.66      & S0A       \\
 NGC 4151                    &  0.0033   &    Sy1.5               & 23.34        &  0.59     & SAB       \\
 IGR J12107+3822       & 0.0229    &    Sy1.5                & 22.67       & 0.82       & E      \\
 IGR J12131+0700      &  0.2095   &    Sy1.5-1.8         &  {\bf 20.14}  &  --         &  --     \\
 NGC4235                    &  0.0080   &    Sy1.2                 & 21.20          &  0.28    & SA   \\
 Mrk 766                       &  0.0129   &    NLS1                 &   $<$21.95   & 0.84     & SB        \\
 NGC 4258                  &  0.0015    &   Sy2                     &  23.03          &  0.49   &  SAB  \\
 PKS 1217+02              & 0.2402    &  Sy1.2                   &  {\bf 20.25}   &  --       & --     \\
 Mrk 50                         &  0.0234   &    Sy1                     &  $<$21.08   &   0.86    & S0       \\
 NGC 4388                   &  0.0084   &    Sy2                    & 23.44           &   0.33   & SAB      \\
 NGC 4395                  & 0.0011    &    Sy2                     & 22.72           &  0.63     & SB     \\
 3C 273                       &  0.1583   &    Sy1/QSO           &  {\bf 20.23}    &  1.00      &  --         \\
 Mrk 771                      &  0.0630    &   Sy1                   &  {\bf 20.88}     & 0.82       &  S0B        \\
 XSS J12303-4232      & 0.1000     &   Sy1.5                & {\bf 20.88}      &  --           &   --       \\
 NGC 4507                  &  0.0118   &    Sy2                   &  23.64             &  0.84       & SAB        \\
 SWIFT J1238.6+0928  & 0.0829  &    Sy2                    & 23.52            & 0.68        & S0        \\
 ESO 506-G27             &  0.0250   &    Sy2                   &  23.03            &  0.44       & S0       \\
 LEDA 170194              &  0.0367  &    Sy2                   & 22.46               & 0.88      & S0       \\
 NGC 4593                   &  0.0090  &    Sy1                   & {\bf 20.30}       &  0.54       & SB        \\
 IGR J12415-5750          &  0.0242       &    Sy1.5        &  {\bf 21.48}        &   0.74    &  S     \\
 IGR J12482-5828          &  0.0280       &    Sy1.9         & 22.35                &  0.60       &  --     \\
 NGC 4748                      &  0.0146       &    NLS1        & {\bf 20.56}         &  0.72       & SA     \\
 ESO 323-32                   &  0.0160       &    Sy2           & 25.00                & 0.74        & S0B    \\
 Mrk 783                          &  0.0672       &    NLS1        &  21.80               & 1.00        &   E      \\
 IGR J13038+5348         &  0.0300       &    Sy1.2        &   {\bf 20.22}        & 0.56       &  --     \\
 NGC 4941                      &  0.0037       &    Sy2          &  23.64                 & 0.53       & SAB     \\
 IGR J13042-1020          &  0.0104       &    Sy2          &  $>$25.00            & 0.71       & SA     \\
 NGC 4945                      &  0.0019       &    Sy2          &  24.72                 & 0.31       & SB   \\
 ESO 323-77                   &  0.0150       &    Sy1.2       &  {\bf 21.57}          & 0.84       &  SB       \\
 IGR J13091+1137         &  0.0291       &    XBONG   &  23.63                 & 0.58         & SB     \\
 IGR J13107-5626          &   --               &   AGN         &  23.59                 &  0.50        & --    \\
 IGR J13109-5552          &  0.104         &    Sy1          & $<$21.66            & --             &  --      \\
 IGR J13133-1109          &  0.0343       &  Sy1            &  {\bf 20.42}          & 0.80        &  --      \\
 IGR J13149+4422         &  0.0353      &   Sy2/LINER  & 22.72               & 0.56          &  S0     \\
 IGR J13168-7157          & 0.0705       &   Sy1.5        &   {\bf 21.21}         & 0.74         &  --        \\
 NGC 5100                      & 0.0319      &   LINER       &  23.16                 & --         & S0       \\
 MCG-03-34-063            &  0.0213       &    Sy2          &  23.59                 &  0.36        &  SB     \\
 Cen A                            &  0.0018       &    Sy2          &  23.17                 &  0.96         & S0      \\
 ESO 383-18                  &  0.0124       &    Sy2          &  23.29                 & 0.50          & S      \\
 MCG-06-30-015            &  0.0077       &    Sy1.2       & 22.17                  & 0.54          &  S       \\
 NGC 5252                     &  0.0230       &    Sy1.9       & 22.83                   & 0.50          & S0     \\
 IGR J13415+3033        &  0.0398       &    Sy2          &  23.47                   &  0.62         & SB     \\  
 SWIFT J1344.7+1934  &  0.0271       &  Sy2/LINER  & 23.39                   & 0.38         & S0      \\
 IGR J13466+1921        &  0.0850       &    Sy1.2         &  {\bf 20.27}          &  0.82        &  S      \\    
 Cen B                            &  0.0129       &    RG/type2   & 22.11                 & 0.80          &   E     \\
 4U 1344-60                   &  0.0130       &    Sy1.5         & 23.67                  &  --             &  S      \\
 IGR J13477-4210         &  0.0386       &    Sy2            & 22.79                  &  0.72         &  --    \\
 IC 4329A                       &  0.0160       &    Sy1.2         & 21.54                 & 0.34          & S0A     \\
 1AXG J135417-3746    &  0.0509       &    Sy1.9         & 22.80                 & 0.78           & --            \\        
\hline
\end{tabular}
\end{table*}

\begin{table*}
\begin{tabular}{|| l c l l l l | c }
  \hline\hline
{\bf Name  }    & {\bf $z$ }  &    {\bf Class}  & Log N$_{H}$ & {\bf b/a}  & Morph/BAR   \\
\hline
 IGR J13550-7218          &  0.0710       &    Sy2           & 23.28         &  --              & --    \\   
 PKS 1355-416               &  0.3130       &    Sy1           & 21.43         &  --              &  --      \\
 IGR J14080-3023          &  0.0237       &    Sy1.5       & {\bf 20.56}  & 0.94           &  --      \\ 
 SWIFT J1410.9-4229    &  0.0339       &  Sy2             &  22.76        & 0.56           & --      \\
 Circinus Galaxy             &  0.0014      &   Sy2            & 25.60         & 0.41            & SA     \\
 NGC 5506                     &  0.0062       &    Sy2           & 22.53         & 0.28           & SAB     \\
 IGR J14175-4641         &  0.0760        &    Sy2           & 23.88         & 0.66           &  --      \\
 NGC 5548                     &  0.0172       &  Sy1.5          & {\bf 20.19}   &  0.86         & S0A      \\
 ESO 511-G030             &  0.0224       &    Sy1           & {\bf 20.70}   &  0.90          & SA     \\ 
 H 1419+480                  &  0.0723       &    Sy1.5        &  {\bf 20.83}  &   --            &  --    \\
 IGR J14301-4158         &  0.0039       &    Sy2           &  22.08         &  0.46          & --     \\
 NGC 5643                     &  0.0040      &    Sy2            & 23.85          &  0.98          & SAB   \\
 NGC 5674                     & 0.0249       &  Sy1.9          & 23.05           &  0.82          & SAB \\
 SWIFT J1436.8-1615    & 0.1445       & Sy1/QSO     & {\bf 20.88}   &  0.88          &  --      \\
 NGC 5728                     &  0.0093      &    Sy2            &  24.14         &  0.44           & SAB   \\
 IGR J14471-6414          &  0.0530     &    Sy1.2          &  21.60        &  --                &  --    \\  
 IGR J14471-6319          &  0.0380     &    Sy2             &  23.39        &   --              &    --     \\
 IGR J14488-4008          &  0.1230     &  Sy1.2            & 22.89         & --                &  --       \\
 IGR J14492-5535          &  --              &  AGN             & 23.08         &  --                & --     \\
 IGR J14515-5542          &  0.0180     &  Sy2               &  21.52        &  0.50           & S     \\
 PKS 1451-375               &  0.3140     &  Sy1.2            & {\bf 20.80}  &   --              & S0       \\
 IGR J14552-5133          &  0.0160     &    NLS1          & {\bf 21.53}  &   0.80          &  S      \\
 IGR J14561-3738          &  0.0240     &    Sy2             & $>$24.00   & 0.66            &  SB    \\
 IC 4518A                       &  0.0163      &    Sy2            & 23.15          &  --               & S       \\
 MKN 841                       &  0.0364      &    Sy1.5         & {\bf 20.84}   &  0.98          &  E       \\
 IRAS 15091-2107         &  0.0446      &    NLS1          & 21.15          & 0.78           &  S     \\
 SWIFT J1513.8-8125    &  0.0684     &   Sy1.2           & 21.88          &  0.60          &  --       \\
 ESO 328-36                  &  0.0237     &   Sy1.8           & {\bf 20.84}  &  --               & S0       \\
 IGR J15301-3840         &  0.0155     &   Sy2              & 22.99         & 0.40             & SA    \\
 IGR J15311-3737          &  0.1270     &  Sy1               & {\bf 21.32}  &    --              &  --        \\
 MCG -01-40-001            &  0.0227     & Sy2               & 22.59          &  0.34           &   S        \\
 IGR J15359-5750          &  --              & AGN             & 23.30          &   --                &   --       \\
 IGR J15415-5029          &  0.0320     &  Sy2?            & $<$22.04    & 0.52            & S      \\  
 NGC 5995                     &  0.0252      &    Sy1.9         & 21.93         &   0.88           & SAB       \\
 IGR J15539-6142          &  0.0150     &    Sy2            &  23.24         &  0.26            & S       \\
 IGR J15549-3740          &  0.0190     &    Sy2            & 22.76          &  0.64            & S        \\
 IGR J16024-6107          &  0.0110     &    Sy2            & 21.40          &  --                 & S        \\
 IGR J16056-6110          &  0.0520     &    Sy1.5         & {\bf 21.31}   & 0.72             &  S      \\
 IGR J16058-7253(1)     &  0.0690     &    Sy2?           & 23.24          & 0.80             & --           \\  
 IGR J16058-7253(2)     &  0.0900     &    Sy2             & 23.58          &  0.52            & --        \\
 IGR J16119-6036         &  0.0160     &    Sy1.5           & {\bf 21.36}  &   0.96           &  SB      \\
 IGR J16185-5928         &  0.0350     &    NLS1           & 23.02          &   0.88          &  SAB        \\
 Mrk 1498                       &  0.0547     &   Sy1.9           &  23.76          & 0.82            &  E             \\
 SWIFT J1630.5+3925   &  0.0306    &   Sy2               &  23.60         & 0.44             & S0           \\
 IGR J16351-5806          &  0.0091    &    Sy2              & 24.68          &  0.87            & SAB         \\
 IGR J16385-2057          &  0.0269    &    NLS1           & {\bf 21.08}   &  0.80            & E        \\
 IGR J16426+6536         &  0.3230    &    NLS1           & {\bf 20.41}   &  --                 & --      \\
 IGR J16482-3036          &  0.0313    &    Sy1              & 21.00          &   0.72            & --       \\
 SWIFT J1650.5+0434   &  0.0321    &    Sy2              & 22.68          &  0.74              & E          \\  
 ESO 138-1                    &  0.0091    &    Sy2               & $>$24.16    &  0.62             & SAB       \\  
 NGC 6221                     &  0.0050    &    Sy2               &  22.04         &  0.74             &  SB       \\   
 NGC 6240                    &  0.0245      &    Sy2/LINER   &  24.30        &  0.56              & S0       \\
 IGR J16558-5203         &  0.0540      &    Sy1.2             &  23.48       & --                   & --     \\
 IGR J17009+3559        &  0.1130      &    XBONG         &  23.37        & 0.52              & --           \\  
 IGR J17036+3734        &  0.0650      &    Sy1                & {\bf 20.39} & 0.84              & E        \\   
 IGR J17111+3910        &  --               &   AGN               & 20.55          &  0.58            & SB     \\  
 NGC 6300                    &  0.0037      &   Sy2                 & 23.38          &  0.67           & SB    \\
 MCG+08-31-041          &  0.0242      & Sy1/LINER       & $<$21.22    &  0.94            & E      \\
 IGR J17204-3554         &  --              &  AGN                & 23.20          &  --                 & --     \\
 Mrk 506                        &  0.0430      & Sy1.5                & $<$20.70    & 0.78            &  SAB  \\
 SWIFT J1723.5+3630  &  0.0400      &  Sy1.5              & 21.28           & 0.46            &  E    \\
\hline
\end{tabular}
\end{table*}

\begin{table*}
 \begin{tabular}{|| l c l l l l | c }
  \hline\hline
{\bf Name  }    & {\bf $z$ }  &    {\bf Class}   & Log N$_{H}$                & {\bf b/a} & Morph/BAR  \\
\hline
 IGR J17348-2045         & 0.0440       &  Sy2                 & 23.23         & --            &  --        \\
 IGR J17379-5957         &  0.0170      &  Sy2                 & {\bf 20.79}  &  0.78      & SA      \\
 GRS 1734-294             &  0.0214      &    Sy1                & $>$21.32  &   --           &  --    \\ 
 2E 1739.1-1210            &  0.0370      &    Sy1.2             & 21.18        & --            &   --       \\
 IGR J17476-2253         &  0.0463     &    Sy1                 & {\bf 21.48} &   --          &  --        \\
 IGR J17488-2338         &  0.2400     &  Sy1.5                & 22.06        &  --           &   --      \\
 IGR J17488-3253         &  0.0200     &  Sy1                   & 21.53        &   --          &  --        \\
 IGR J17513-2011         &  0.0470     &    Sy1.9              & 21.82        &  --           & --          \\
 IGR J17520-6018         &  0.1120     &  Sy2                   & 23.11        & 1.00        &   --       \\
1RXS J175252.0-053210   &  0.1360    &    Sy1.2         & {\bf 21.33} & --            &  --          \\
 NGC 6552                     &  0.0265     & Sy2                   &  23.84        & 0.46       &  SB    \\        
 IGR J18027-1455          &  0.0350     & Sy1                   & 21.48        &   --          &   --         \\
 IGR J18078+1123         & 0.0780      &   Sy1.2              & 22.25        &  0.92       &  --        \\
 IGR J18129-0649          & 0.7750       &  Sy1/QSO       & {\bf 21.54} &  --            &  --         \\
 SWIFT J1821.6+5953   & 0.0990       &  Sy2                 &  22.88       & 1.00        &  --   \\
 IGR J18218+6421         &  0.2970      &    Sy1.2            & {\bf 20.54} &  --           &  --      \\  
 IGR J18244-5622          &  0.0169     &    Sy2                & 23.15        & 0.38        &  SB     \\
 IGR J18249-3243          &  0.3550     &    Sy1/QSO      & {\bf 21.14} &    --          &   --      \\   
 IGR J18259-0706          &  0.0370     &   Sy1                 &  21.91       &   --           &  --   \\     
 IGR J18308+0928         &  0.0190     &  Sy2                  &  23.07       & 0.56        &  --   \\
 IGR J18311-3337          &  0.0687    &   Sy2                  & 22.15        & 0.74        &   --   \\
 3C 382                           &  0.0579     &   Sy1                 & {\bf 20.79} & 0.68       &   E      \\
 Fairall 49                       &  0.0200     &  Sy2                    & 22.09       &  0.88       & S    \\
 ESO 103-35                 &  0.0133    &   Sy2                     & 23.30       & 0.36        & S0A     \\
 3C 390.3                       &  0.0561    &   Sy1.5               & {\bf 20.63}  &  0.84       & E    \\
 ESO 140-43                 &  0.0142     &   Sy1.5                 & 23.04      &   0.44       & SB    \\
 SWIFT J1845.4+7211  &  0.0463   &  Sy2                      & 22.51      &  0.42         & S       \\
 IGR J18470-7831         &  0.0743   &   Sy1                     &  20.08     &  0.60         &  --      \\    
 PBC J1850.7-1658       &   --           &  AGN                    &  22.17     &  --             & --    \\
 IGR J18538-0102         & 0.1450    & Sy1                       & 21.61       & --            &  --   \\
 ESO 25-2                      &  0.0292    &   Sy1                    & {\bf 20.92} &  0.98     & SB    \\
 2E 1853.7+1534            &  0.0840   &   Sy1                   &  $<$21.59   &   0.92   &   --    \\
 2E 1849.2-7832             &  0.0420    &   Sy1                   & {\bf 20.92}  &  0.56     &  S0B   \\
 IGR J19077-3925           &  0.0760     &   Sy1.9              &  21.15        &  0.58      &  --   \\
 IGR J19118-1707          &  0.0234    &   LINER               & 23.02         &  0.72      & S    \\
 PKS 1916-300              &  0.1668    &   Sy1.5-1.8          & {\bf 20.90}   &     --       &    -- \\
 ESO 141-G055              & 0.0371   &   Sy1.2                  & {\bf 20.68}  &  0.82     & S      \\ 
 SWIFT J1930.5+3414    &  0.0629    &   Sy1.5-1.8         & 23.44          &  --           & --   \\
 SWIFT J1933.9+3258       & 0.05775    & Sy1.2                & $<$20.60     & --        & --   \\
 IGR J19378-0617          &  0.0106    &   NLS1               & 23.50           &  0.86      &  E     \\
 IGR J19405-3016           &  0.0520   &   Sy1.2               & {\bf 20.96}   & 0.70       & --     \\
 NGC 6814                     &  0.0052    &   Sy1.5                & {\bf 21.10}  &    0.94    & SAB   \\
 XSS J19459+4508        &  0.0539    &   Sy2                  & 23.04       &  0.56         & --   \\    
 IGR J19491-1035          &  0.0246    &   Sy1.2               & 22.26       &   0.82        &  E     \\ 
 3C 403                           &  0.0590     &    Sy2                & 23.65       &   0.74        & E   \\
 Cyg A                            &  0.0561     &    Sy2                &  23.30       & 0.84            & S   \\
 ESO 399-20                  &  0.0249     &    NLS1              & {\bf 20.85} & 0.64           &  E   \\
 NGC 6860                     &  0.0149     &    Sy1.5              & 21.00       &  0.64          &  SB  \\ 
 SWIFT J2018.4-5539    &  0.0606    &   Sy2                  &  23.50      &  0.60            &  E   \\
 IGR J20186+4043         &  0.0144    &    Sy2                 & 22.76      &   0.74            & --     \\   
 IGR J20216+4359          &  0.0170   &    Sy2                & 23.11      &   0.68             & --      \\   
 IGR J20286+2544(1)     &  0.0139    &    Sy2/SB          & 23.78      &  0.44              & S \\
 IGR J20286+2544(2)     &  0.0144    &   Sy2                & 23.97      &   0.40              & S0A  \\
 NGC 6926                      &  0.0196    &  Sy2                  & 24.00      & 0.46               & SB     \\
 4C +21.55                      &  0.1735    &   Sy1                  &  21.78     & 0.36             &  --    \\
 4C 74.26                        &  0.1040    &    Sy1                 & 21.15          &   --             & --        \\
 SWIFT J2044.0+2832   & 0.0500      &    Sy1                &{\bf 21.24}  &  --                 & --    \\    
 Mrk 509                          & 0.0344     &    Sy1.5              & {\bf 20.63}  &   0.82         & --     \\
 IGR J20450+7530         & 0.0950      &   Sy1                 & 21.20          & 1.00           &  --  \\ 
 S52116+81                    &  0.0840     &    Sy1                & {\bf 21.38}   &   0.70        &  --  \\
 IGR J21178+5139         &   --            &  AGN                 &  22.32   & --                      & --    \\
\hline
\end{tabular}
\end{table*}

\begin{table*}
 \begin{tabular}{|| l c l l l l | c }
  \hline\hline
{\bf Name  }    & {\bf $z$ }  &    {\bf Class}   & Log N$_{H}$                  & {\bf b/a} & Morph/BAR  \\
\hline
 1RXS J211928.4+333259   &  0.0510   &    Sy1.5         &  21.52          & --    &    --        \\
 IGR J21247+5058               &  0.0200     &    Sy1               &  22.89         &     --     &  --     \\
 SWIFT J2127.4+5654         &  0.0147    &    NLS1             & {\bf 21.90}   &   --       &   --     \\
 CTS 109                           & 0.0299     &  Sy1.2              &  {\bf 20.55}  &  0.80  &  S0     \\
 RX J2135.9+4728           &  0.0250     &    Sy1               &  20.30         &  --       &  --     \\
 1RXS J213944.3+595016     &  0.1140     &    Sy1.5     & 21.54           & --         &   --      \\
 IGR J21565+5948           & 0.2080      &    Sy1              &  {\bf 21.81}   &     --    &    --      \\   
 SWIFT J2156.2+1724     &  0.0340     &  Sy1.8              & 23.33           &  1.00   & --      \\
 Mrk 520                           &  0.0266      &    Sy1.9            & 22.63          & --     & S0      \\
 NGC 7172                      &  0.0087      &    Sy2                & 22.86          &  0.50    & SA  \\
 UGC 12040                    &  0.0213      &  Sy1.9             &  $<$ 22.92   & 0.66      & S0B     \\
 IGR J22292+6647           &  0.1120      &    Sy1.5         &{\bf 21.68}     & --            &  --       \\    
 NGC 7314                       &  0.0048     &    Sy1.9          & 22.02            &  0.44      & SAB     \\
 IGR J22367-1231           & 0.0241      &    Sy1.9           &  22.42           &  0.44      & SA       \\        
 3C 452                            &  0.0811     &    Sy2              &  23.77          &  0.66       & E      \\
 QSO B2251-178            &  0.0640     &    Sy1.2            & 22.33           &  0.84      & S0      \\     
 KAZ 320                         & 0.0345      &   NLS1             & {\bf 20.68}    &  0.72     & S        \\
 NGC 7465                      &  0.0066     &    Sy2/LINER   &  23.66          &  0.74       & SB     \\
 NGC 7469                      &  0.0163     &    Sy1.5            & {\bf 20.46}    &  0.91     &  SAB     \\
 MCG-02-58-022             &  0.0469     &    Sy1.5            & {\bf 20.56}    &  0.86      &  SA     \\     
 NGC 7479                      &  0.0079     &  Sy1.9              &  24.30          &  0.40      & SB   \\
 NGC 7582                       &  0.0052     &    Sy2              &  24.04          & 0.35       & SB     \\
 IGR J23206+6431          &  0.0717     &    Sy1               &  21.95         & 0.70        &   --      \\  
 RHS 61                          & 0.1200      &   Sy1                &  {\bf 20.59}  & 1.00       &   --     \\    
 IGR J23308+7120          &  0.037      &    Sy2 ?             &  22.96          &   0.92     & --    \\     
 IGR J23524+5842          &  0.164      &    Sy2 ?             & 22.46             &     --      &  --     \\
 IGR J23558-1047           &  1.1080    &   Sy1/QSO       & {\bf 20.41}      &   --        &   --    \\    
\hline
\end{tabular}
\begin{tablenotes}
\item Note: N$_{H}$ values in bold are referring to upper limits or Galactic values.
\end{tablenotes}
\label{all}
\end{table*}

\end{center}

\clearpage

\begin{center}
\begin{table}
\begin{sideways}
\begin{threeparttable}[b]
\caption[]{INTEGRAL/IBIS  AGN}
 \begin{tabular}{l c c  c l l l c  l  l l l  }
 \hline\hline
  \multicolumn{12}{c}{\bf {Mereminsky}} \\
  \hline\hline
 Name	                                &  RA                 & dec                   &  z          & class           & Log N$_{H}$   & $\Gamma_{2-10~keV}$&       F$_{S}^{\dagger}$&  F$_{H}^{\ddagger}$ & b/a    & Morph/BAR & 	Ref.	\\
 \hline\hline
SWIFT J0308.5-7251               & 03 07 35.32   & $-$72 50 02.5  & 0.0276  &  Sy1.2         &  --                   & 1.80$\pm$0.17      & 4.80                    & 7.56                      & 0.80      & S                 & 1 \\
SWIFT J0422.7-5611               & 04 22 24.18    & $-$56 13 33.5  & 0.0435  &  Sy2            & 22.89             & 1.38$\pm$0.21      & 3.40                    & 16.00                    &  --         & S                 & 1 \\
1H0419-577$^{(a)}$                 & 04 26 00.72   & $-$57 12 01.0   & 0.1040   & Sy1.5         &19-22             & 1.70$\pm$0.01      & 8-16                   & 22.34                     &  --         &  --                & 2 \\
SWIFT J0440.2-5941               & 04 38 59.00  & $-$59 40 54.0   & 0.0577   & Sy2             & 22.34            & 2.12$\pm$0.27       & 2.80                  & 7.74                       &  0.38     & --                 & 1\\
SWIFT J0504.6-7345               & 05 04 34.20  & $-$73 49 27.1   & 0.0452   & Sy1.9          & 21.91            & 1.78$\pm$0.15       & 3.40                  & 8.44                       & 0.62       & --                & 1 \\
SWIFT J0505.6-6735               & 05 05 24.35  & $-$67 34 35.4     & 0.046  & likely Sy2     & 23.74           &  1.50$\pm$0.76      & 1.00                 & 10.90                       & 0.42       & --                & 1\\
IGR J05347-6015                     & 05 34 30.90  & $-$60 16 15.6   & 0.0570    & Sy1             & 22.56          & 1.96$\pm$0.50        & 4.73                 & 15.13                      & 0.48      & --                 & 3 \\
SWIFT J0634.7-7445               & 06 34 03.52   & $-$74 46 37.6  & 0.1120     & Sy1            &  --                & 1.61$\pm$0.25        & 2.10                 &  7.56                       & --           & --                 & 1 \\
IGR J06569-6534                     & 06 56 29.20   & $-$65 33 38.0  & 0.0295    & Sy1             & $<$22.23    &  1.77$\pm$0.50      & 1.70                 &  9.00                        & 0.78      &  S               & 3 \\
SWIFT J0747.6-7326               & 07 47 38.36   & $-$73 25 53.2  & 0.0360    & Sy2              & 23.56         & 1.40$\pm$0.53       & 1.30                 &  12.31                      & 0.92      &  E                & 1 \\
NGC 2655                                & 08 55 37.73    & $+$78 13 23.1  & 0.0047  & LINER          &  23.36        &  1.9 (fixed)              &  1.05                 &  7.56                       &  0.90     & SAB               &  4 \\
SWIFT J0935.9+6120              & 09 35 51.60    & $+$61 21 11.4  & 0.0394  & Sy2              & 24.12        &  1.63$\pm$0.20        &  0.20                 & 5.10                        &  0.70    &   S0               &   5 \\
SWIFT J1033.6+7303              & 10 34 23.54    & $+$73 00 50.2  & 0.0220  & XBONG      & 22.62          &  1.53$\pm$0.22       &  2.60                & 7.21                          & 0.48     & S0              &  1 \\
SWIFT J1044.1+7024              & 10 44 08.54   & $+$70 24 19.3  & 0.0336    & Sy2            & 23.27          &  1.58$\pm$0.35        &  2.20               & 8.97                          & 0.36      & S           & 1\\
SWIFT J1105.7+5854$^{(b)}$ & 11 06 49.55   & $+$57 41 07.7  & 0.0322    & Sy2            & 23.87           & 2.20 (fixed)                & 0.66                & 11.78                       & 0.60       &  S0           &  6\\
SWIFT J1114.3+7944              & 11 14 43.91   & $+$79 43 35.9  & 0.0372   & Sy2             & 22.66           & 1.42$\pm$0.67          & 1.70                & 7.39                        & 0.56        &  --             & 1\\
SWIFT J1144.1+3652              & 11 44 29.88   & $+$36 53 08.6 & 0.03801   & Sy1            & --                  & 1.80$\pm$0.04         & 1.84             & 9.85                          & 0.90         & S             & 6 \\
SWIFT J1148.3+0901              & 11 47 55.08   & $+$09 02 28.8  & 0.0688    & Sy1.5          &--                  & 1.78$\pm$0.08         & 4.00             & 6.16                          & 0.74         &   --             & 6 \\
SWIFT J1148.7+2941              & 11 48 45.95    & $+$29 38 28.3  & 0.0230    & LINER       &  22.70           & 1.51$\pm$0.81        & 5.15             & 12.66                        & 0.48         & S0              & 6 \\
SWIFT J1201.2-0341               & 12 01 14.36    & $-$03 40 41.1  &  0.0196    & Sy 1          & --                   & 1.98$\pm$0.15       & 6.10              & 10.20                        & 0.80         & E               & 1 \\
3PBC J1204.7+3109                & 12 04 43.32    & $+$31 10 38.2 & 0.0250    & Sy1.9         & 23.00             & 1.88$\pm$0.08       & 2.30             &  9.67                       & 0.90          & SB              & 6 \\
SWIFT J1207.5+3355              & 12 07 32.90    & $+$33 52 40.0 & 0.0791   & Sy2            & 22.70              & 1.50$\pm$0.13      & 1.60              & 7.74                        & 0.68           & E                & 1 \\
NGC 4180                                & 12 13 03.05    & $+$07 02 20.2 & 0.0070    & LINER       & 24.16              & 1.92$\pm$0.12       & 0.20              & 23.39                      & 0.38           & SB           &  1\\
WAS 49B                                 & 12  14  17.73  & $+$29 31 43.2 & 0.0640    & Sy2           & 23.41               & 2.03$\pm$0.33       & 2.60             & 9.85                          & 0.96        &  --                &   1\\
IGR J12171+7047                    & 12 17 26.20   & $+$70 48 07.2  & 0.0067  & AGN            & 23.40             & 1.7 (fixed)                & 0.34            & 8.44                         & 0.50          & S0B               & 3\\
MRK 205                                   & 12 21 44.22   & $+$75 18 38.8  & 0.0708  & Sy1             & 23.71            &  1.97$\pm$0.02       & 0.93             & 13.54                       & --              & SB             & 7\\
3PBC J1231.3+7044                & 12 31 36.44   & $+$70 44 14.1  & 0.2080  & Sy1.2          & 22.78             & 2.01$\pm$0.07       & 0.57              & 8.62                         & --              &  --                 & 8\\
NGC 4579                               & 12 37 43.52     & $+$11 49 05.5 & 0.0051    & Sy1.9        &  --                   & 1.78$\pm$0.1         & 3.91               & 8.79                        & 0.95          & SAB                & 3\\
SWIFT J1240.9+2735              & 12 40 46.41     & $+$27 33 53.7 & 0.0567    & Sy2           & 22.45             & 1.23$\pm$0.45      & 2.10               & 8.79                       & 0.78           & S             & 1\\
NGC 4736                               & 12 50 53.06     & $+$41 07 13.6  & 0.0010    & Sy2/LINER & 20.90            & 2.00$\pm$0.15       & 0.14              & 8.79                       & 0.85           & SA              & 9 \\
SWIFT J1300.1+1635             & 13 00 05.35     & $+$16 32 14.9 & 0.0800     & Sy2             & 22.5             & 1.50$\pm$0.17        & 2.04             & 6.16                       & 0.78            &  --                & 5 \\
NGC 5033                               & 13 13 27.47     & $+$36 35 38.2  & 0.0029    & Sy1.9          & --                 & 1.70$\pm$0.1         & 2.87              & 11.26                     & 0.56             & --                 & 10\\
3PBC J1342.0+3539               & 13 42 08.34   & $+$35 39 15.2  & 0.0036     & Sy1.9          & 21.95          & 1.40$\pm$0.1           & 6.71            &  17.76                   & 0.84               & S0A               &  10 \\
\hline
\hline
\end{tabular}
\begin{tablenotes}
      \item   $^{(\dagger)}$: 2-10 keV flux in units of 10$^{-12}$ erg cm$^{-2}$ s$^{-1}$
      \item   $^{(\ddagger)}$: 20-100 keV flux in units of 10$^{-12}$ erg cm$^{-2}$ s$^{-1}$
             \item (a): highly variable source both in flux and spectral shape (see Di Ges\'u et al. 2014)
           \item (b): blended source Sy2 (CGCG 291-028) and QSO, in table values referred to the Sy2 only (not included in the analysis)  \\            
                                    References: (1): \citealt{Ricci_2017} ;(2): \citealt{Di_Gesu_2014};  (3): {\bf this work}; (4): \citealt{Brightman_2011};  (5): \citealt{Oda_2017}; (6): \citealt{Vasudevan_2013};
                                                       (7):\citealt{Patrick_2012};  (8): \citealt{Miniutti_2010}; (9): \citealt{Gonz_lez_Mart_n_2006}; (10): \citealt{Cappi_2006}; 
\end{tablenotes}
\end{threeparttable}
\end{sideways}
\label{mere}
\end{table} 
\end{center}

\clearpage

\begin{center}
\begin{table}
\begin{sideways}
\begin{threeparttable}[b]
\caption[]{INTEGRAL/IBIS  AGN}
 \begin{tabular}{l c c  c l l l c  l  l l l }
 \hline\hline
  \multicolumn{12}{c}{\bf {Krivonos}} \\
  \hline\hline
 Name	                    &  RA                 & dec                   &  z          & class        & Log N$_{H}$  & $\Gamma_{2-10~keV}$ &       F$_{S}^{\dagger}$&  F$_{H}^{\dagger}$ & b/a    & Morph/BAR & 	Ref.	\\
 \hline\hline
IGR J03117$+$5028   & 03 11 54.72      & $+$50 30 21.1  & 0.0624   & Sy1.5     & 21.08              & 1.80$\pm$0.16     &  5.9                &  8.64                      &  0.56   &    --                        &    1 \\ 
IGR J07396$-$3143    & 07 39 44.69      & $-$31 43 02.5   & 0.0261  & Sy2         & 23.46             &  1.41$\pm$0.31    &  1.0                 &  21.35                   &  0.48    &  S0                        &    1 \\ 
IGR J07433$-$2544    & 07 43 14.72      & $-$25 45 50.1   & 0.023    & Sy1         & 20.95             &  2.08$\pm$0.20    &  9.5                 &  15.05                   &  0.84    &  --                           &   1 \\ 
IGR J08398$-$1214    & 08 39 50.59      & $-$12 14 34.3   & 0.198    & Sy1         & 20.00             &  2.03$\pm$0.08     &  11.4               &  22.21                   &  --         &  --                          &   1  \\ 
IGR J16494$-$1740    & 16 49 21.02      & $-$17 38 40.3   & 0.0229  & Sy2         & 22.25             &  1.36$\pm$0.36    &   2.3                &  9.87                      & 0.18     &  S                        & 3 \\ 
IGR J19039$+$3344   & 19 03 49.16      & $+$33 50 40.7  & 0.0150  & Sy2         & 22.87             &   1.79$\pm$0.18   &   4.6                 &  12.96                   &  0.40    &  SB                      &  1  \\ 
IGR J19260$+$4136   & 19 26 30.22      & $+$41 33 04.9  & 0.0720   & Sy2        & 20.49             &   1.71$\pm$0.08   &   5.5                 &   8.14                    &  0.52    &   --                         &   1 \\ 
IGR J21099$+$3533   & 21 09 31.88      & $+$35 32 57.6  & 0.2023   & Sy1        & 21.10             &   2.11$\pm$0.07   &   3.5                  &   8.76                    &   0.62    &  --                         &   3 \\ 
IGR J21382$+$3204   & 21 38 33.41      & $+$32 05 06.1  & 0.0251   & Sy1.5     & 22.44             &   1.82$\pm$0.35   &   2.5                  &  15.67                   &   0.56    &  S                       &   3 \\ 
IGR J21397$+$5949  &  21 39 44.97      & $+$59 50 15.0  & 0.114     & Sy1.5     & 21.46             &  1.74$\pm$0.82    &   7.0                  &  9.62                     &  --         &   --                         & 1  \\ 
\hline\hline
\end{tabular}
\begin{tablenotes}
\item   $^{(\dagger)}$: 2-10 keV flux in units of 10$^{-12}$ erg cm$^{-2}$ s$^{-1}$; 
 \item   $^{(\ddagger)}$: 20-100 keV flux in units of 10$^{-12}$ erg cm$^{-2}$ s$^{-1}$\\
 \item References as in Tab A1 \\
\\
\end{tablenotes}
\end{threeparttable}
\end{sideways}
\label{krivo}
\end{table} 
\end{center}

\clearpage




\label{lastpage}
\end{document}